\documentclass{aa}  

\usepackage{graphicx}
\usepackage{txfonts}
\usepackage[colorlinks,citecolor=blue]{hyperref}
\begin{document}

\title{Dissipation of the striped pulsar wind}

%\subtitle{}

\author{B. Cerutti \inst{1}\and A.~A. Philippov\inst{2,3}\thanks{Einstein Fellow}}

\institute{Univ. Grenoble Alpes, CNRS, IPAG, 38000 Grenoble, France\\
           \email{benoit.cerutti@univ-grenoble-alpes.fr}
           \and
           Department of Astrophysical Sciences, Princeton University, Princeton, NJ 08544, USA
           \and
           Department of Astronomy, University of California Berkeley, Berkeley, CA 94720-3411, USA\\
           \email{sashaph@berkeley.edu}
           }

\date{Received \today; accepted \today}

% \abstract{}{}{}{}{} 
% 5 {} token are mandatory
 
\abstract
{Rapidly rotating neutron stars blow a relativistic, magnetized wind mainly composed of electron-positron pairs. The free expansion of the wind terminates far from the neutron star where a weakly magnetized pulsar wind nebula forms, implying efficient magnetic dissipation somewhere upstream.}
{The wind current sheet that separates the two magnetic polarities is usually considered as the most natural place for magnetic dissipation via relativistic reconnection, but its efficiency remains an open question. Here, the goal of this work is to revisit this issue in light of the most recent progress in the understanding of reconnection and pulsar electrodynamics.}
{We perform large two-dimensional particle-in-cell simulations of the oblique rotator to capture the multi-scale evolution of the wind. Our simulations are limited to the equatorial plane.}
{We find that the current sheet breaks up into a dynamical chain of magnetic islands separated by secondary thin current sheets. The sheet thickness increases linearly with radius while the Poynting flux decreases monotonically as reconnection proceeds. The radius of complete annihilation of the stripes is given by the plasma multiplicity parameter at the light cylinder. Current starvation within the sheets does not occur before complete dissipation as long as there is enough charges where the sheets form. Particles are efficiently heated up to a characteristic energy set by the magnetization parameter at the light cylinder. Energetic pulsed synchrotron emission peaks close to the light cylinder, and presents sub-pulse variability associated with the formation of plasmoids in the sheet.}
{This study suggests that the striped component of the wind dissipates far before reaching the termination shock in isolated pulsars, even in very-high-multiplicity systems such as the Crab pulsar. Pulsars in binary systems may provide the best environments to study magnetic dissipation in the wind.}

\keywords{acceleration of particles -- magnetic reconnection -- radiation mechanisms: non-thermal -- methods: numerical -- pulsars: general -- stars: winds, outflows}
               
\maketitle

%-------------------------------------------------------------------

\section{Introduction}

It is commonly accepted that pulsars launch an ultrarelativistic, ultramagnetized wind of electron-positron pairs beyond the light-cylinder radius where co-rotation with the star is impossible \citep{2009ASSL..357..421K, 2012SSRv..173..341A, 2016JPlPh..82e6302P, 2017SSRv..207..111C}. Observations of pulsar wind nebulae indicate that the plasma is purely non-thermal and weakly magnetized downstream of the wind termination shock \citep{1974MNRAS.167....1R, 1984ApJ...283..694K, 1992ApJ...397..187B}. Therefore, there must be an efficient transfer of the magnetic energy into particle kinetic energy somewhere between the pulsar magnetosphere and the nebula. This puzzle is well-known in the pulsar community as the ``sigma problem'', but this is in fact a more generic problem of all relativistic magnetized outflows such as relativistic jets found in active galactic nuclei (AGN) and gamma-ray bursts. It is still unclear how and where magnetic dissipation happens in pulsars, but these objects are ideal testbeds for understanding this physical process.

The wind current sheet that forms between the two predominantly toroidal magnetic polarities is often suspected to be the main place for magnetic dissipation. If the magnetic axis is aligned with the pulsar rotation axis, the sheet is flat and fills the equatorial plane. If the magnetic axis is inclined at an angle $\chi\neq 0$, the sheet is oscillating and fills an equatorial wedge contained within $\pi/2\pm\chi$ and has the shape of a ballerina's skirt of wavelength set by the light-cylinder radius ($2\pi R_{\rm LC}$) that we refer to herein as the ``striped wind'' \citep{1990ApJ...349..538C, 1999A&A...349.1017B}. Using a laminar model of magnetic dissipation, \citet{1990ApJ...349..538C} and \citet{1994ApJ...431..397M} found complete annihilation of the striped wind far before it reaches the termination shock. \citet{2001ApJ...547..437L} revisited this scenario and found that the wind accelerates as dissipation proceeds so that relativistic time dilation effects would not allow the wind to dissipate significantly, unless the particle density is higher than usually expected \citep{2003ApJ...591..366K}. This conclusion lead \citet{2003MNRAS.345..153L} to propose that the stripes survive up to the termination shock where they annihilate, leading to efficient plasma heating and non-thermal particle acceleration \citep{2007A&A...473..683P, 2008ApJ...682.1436L, 2011ApJ...741...39S}.

Regardless of where dissipation happens, the main physical mechanism at the origin of magnetic dissipation is reconnection. In the context of pulsars where the magnetic energy exceeds the rest mass energy of the plasma (high magnetization, $\sigma\gg 1$), reconnection proceeds in the relativistic regime  \citep{1994PhRvL..72..494B, 2003ApJ...589..893L, 2005MNRAS.358..113L}, that is, the plasma entering the reconnecting regions becomes necessarily relativistic. Recent {\em ab-initio} particle-in-cell (PIC) simulations have shown that relativistic reconnection is fast and efficient at dissipating the magnetic energy for both pair plasmas and electron-ion plasmas, and at producing broad hard particle spectra (see \citealt{2015SSRv..191..545K} for a review and references therein). These studies also revealed that fast reconnection is driven by the relativistic tearing instability \citep{1979SvA....23..460Z} which fragments the current sheet into a hierarchical chain of magnetic islands separated by short current sheets \citep{2010PhRvL.105w5002U}. In the close environment of pulsars, radiative losses are extreme (mostly curvature and synchrotron radiation) and effectively decrease the thickness of the layer, thus leading to an enhancement of the reconnection rate \citep{2011PhPl...18d2105U, 2014ApJ...780....3U}. Pair creation at the base of the striped wind {\em via} photon-photon annihilation is another peculiarity of reconnection in pulsars \citep{1996A&A...311..172L}. The goal of this paper is to revisit the issue of magnetic dissipation and particle acceleration in the striped wind in light of the most recent advances in our understanding of relativistic reconnection.

Global PIC simulations of pulsar magnetospheres have already identified the important role of relativistic reconnection in the energy transfer between the fields and the particles \citep{2014ApJ...785L..33P, 2014ApJ...795L..22C, 2015ApJ...801L..19P, 2015MNRAS.448..606C, 2015MNRAS.449.2759B, 2015ApJ...815L..19P}, eventually leading to energetic pulsed emission \citep{2016MNRAS.457.2401C, 2017arXiv170704323P}. Simulations show efficient dissipation close to the light cylinder within $1$-$2R_{\rm LC}$, from $15$-$20\%$ of the pulsar spindown power for an aligned rotator, down to a few percent for an orthogonal rotator \citep{2015ApJ...801L..19P}. But these studies focus on the magnetosphere and hence are limited to a few $R_{\rm LC}$ in radial extent, so the question of whether dissipation occurs at larger radii further in the wind but before the termination shock remains unanswered.

Here, we present the results of large box-size PIC simulations aimed at capturing the large-scale dynamics of the striped wind self-consistently. We pay little attention to the magnetosphere in this work, even though it is present in the simulation (see previous works). To reach large radial extensions, we perform two-dimensional (2D) simulations of the striped wind in the equatorial plane, starting from a split-monopole configuration. We begin with a detailed description of this numerical setup in Sect.~\ref{sect_setup} and present our results in Sect.~\ref{sect_results}, with an emphasis on the dynamics of reconnection along the striped wind. We also investigate how particle acceleration proceeds within the current sheet and we model the high-energy radiative signatures. Finally, we discuss the implications of these results in Sect.~\ref{sect_ccl}.

\section{Numerical approach and setup}\label{sect_setup}

\begin{figure}
\centering
\includegraphics[width=\hsize]{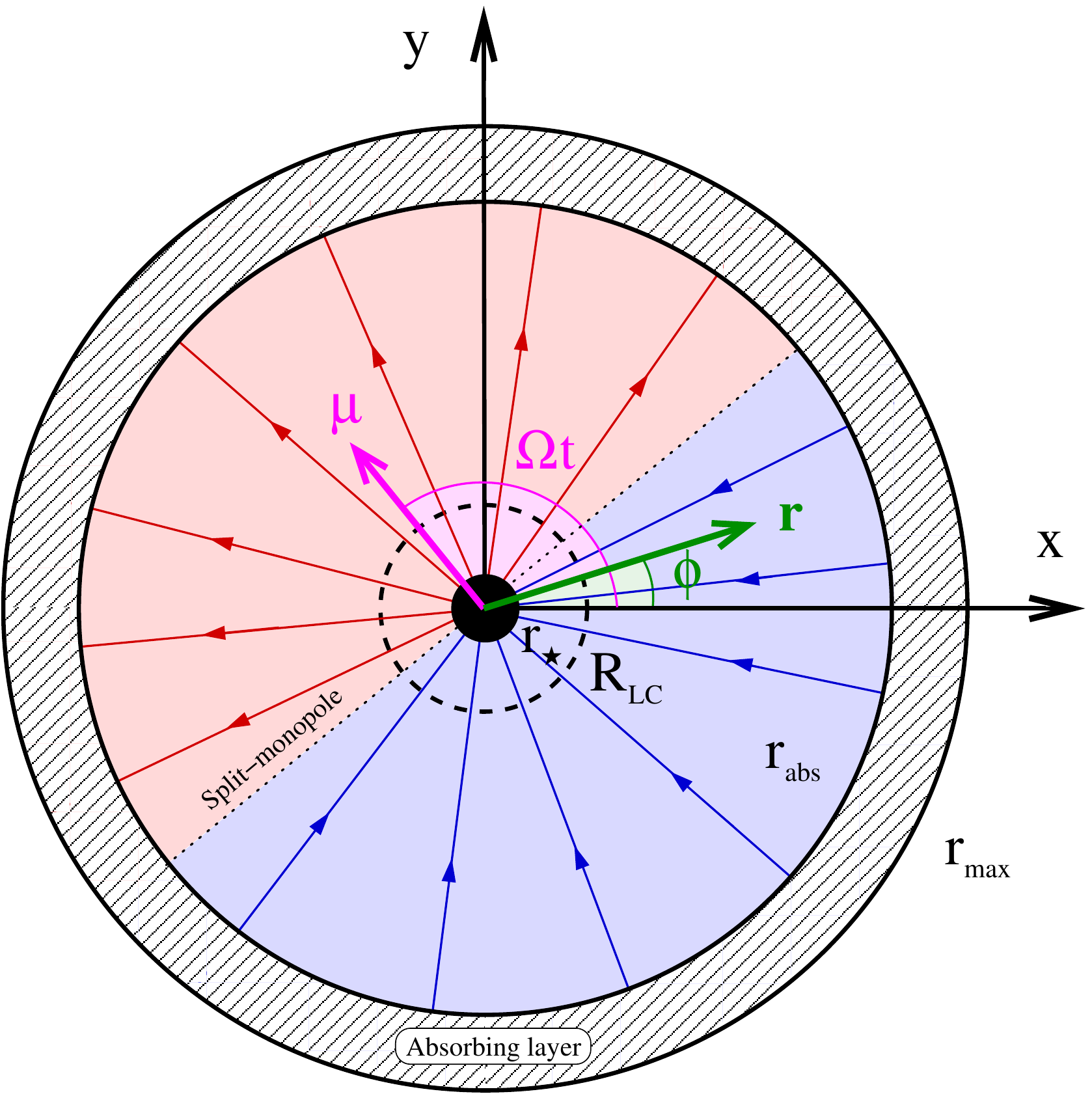}
\caption{Numerical setup of a two-dimensional (2D) PIC simulation of the striped wind in the equator of an inclined split-monopole \citep{1999A&A...349.1017B}. Initially, magnetic field lines are purely radial and their polarity change at $\phi=\Omega t\pm\pi/2$ (dotted line). The box extends from the stellar surface $r_{\rm min}=r_{\star}$ (black disk) where $e^{\pm}$ pairs are created and where the fields are in co-rotation with the star, up to an absorbing layer between $r_{\rm abs}$ and $r_{\rm max}$ where particles and fields vanish. The light-cylinder radius $R_{\rm LC}=c/\Omega$ is shown by the dashed circle.}
\label{fig_setup}
\end{figure}

We use the relativistic electromagnetic PIC code {\tt ZELTRON} \citep{2013ApJ...770..147C} with spherical coordinates ($r,\theta,\phi$) as described in \citet{2015MNRAS.448..606C, 2016MNRAS.457.2401C}. In order to reduce numerical cost and hence capture the large-scale evolution of the striped wind, we restrict the spatial grid to the equatorial plane ($\theta=\pi/2$). The numerical domain is therefore a spherical wedge in the $r\phi$-plane and should not be confused with polar coordinates\footnote{It is worth noting that a relativistic pulsar wind cannot be achieved in a cylindrical geometry, more specifically we find that $E_{z}\neq B_{\phi}$ in this case.}. Time-dependent Maxwell's equations are solved with the usual second-order accurate Yee-algorithm \citep{1966ITAP...14..302Y} directly on the spherical mesh using cell-integrated formulae while particle trajectories are solved in Cartesian coordinates using the Boris push \citep{1991ppcs.book.....B}. Particle positions and velocities are then remapped in spherical coordinates at every time step for current deposition. Particle positions are confined within the $r\phi$-plane but all three components of the 4-velocity vector are retained. In addition to the Lorentz force, particles feel the radiation-reaction force due to the emission of curvature and synchrotron radiation (see \citealt{2016MNRAS.457.2401C} for its implementation), the two main radiative processes within the close environment of pulsars.

The computational domain is a full disk ($\phi\in\left[0,2\pi\right]$) of inner radius $r_{\rm min}$ and outer radius $r_{\rm max}$ (see Fig.~\ref{fig_setup} for an overview of the setup). We place the inner boundary at the surface of the star $r_{\rm min}=r_{\star}$ and the outer boundary far away from the light-cylinder radius, $r_{\rm max}=170 R_{\rm LC}$, where $R_{\rm LC}=3r_{\star}$ in our fiducial simulations. We have also investigated slower rotators with $R_{\rm LC}=6 r_{\star}$ and $9 r_{\star}$. The outer boundary of the box is coated with a $\left(r_{\rm max}-r_{\rm abs}\right)=0.1r_{\rm max}$ thick layer that smoothly absorbs all particles and electromagnetic waves leaving the box. This open boundary can be achieved by adding finite electric and magnetic conductivities within the absorbing layer \citep{2015MNRAS.448..606C, 2015NewA...36...37B}. It implicitly assumes that no information will be able to return inward. We see below that this is a very good assumption in this context because the wind crosses the fast magnetosonic point well before it reaches the outer boundary. The full domain is composed of $\left(4096\times4096\right)$ cells with a logarithmic spacing in radius and a constant spacing in $\phi$. This configuration is well adapted for modeling the striped wind because the fields and the density decrease with radius (as $1/r$ and $1/r^2$, respectively, for $r\gg R_{\rm LC}$) and hence the particle Larmor radius ($\rho_{\rm L}$) and skin-depth ($d_{\rm e}$) increase with radius. It also allows to keep the cell aspect ratio constant with radius ($\Delta r/r\Delta \theta=$constant).

We assume that the neutron star is a perfect conductor rotating at the constant angular velocity $\boldsymbol{\Omega}=(c/R_{\rm LC})\mathbf{e_{\rm z}}$. Magnetic field lines are frozen into the surface so that their solid rotation with the star induce the co-rotation electric field
\begin{equation}
\mathbf{E}_{\star}=-\frac{\left(\boldsymbol{\Omega}\times\mathbf{r}_{\star}\right)\times\mathbf{B}_{\star}}{c},
\end{equation}
at the inner boundary. Initially, the box is in vacuum (no particles) with a static magnetic field configuration. We first tried to create a striped wind from an inclined rotating dipole as it is usually done in the literature, but we found that this choice is not appropriate in this numerical setup because the assumption of a zero-gradient along the $\theta$-direction is not valid for a dipole ($\partial B/\partial\theta\neq 0$): we found unphysical cyclic changes of the magnetic structure, even in vacuum. Instead, we chose a split-monopole configuration \citep{1973ApJ...180L.133M, 1999A&A...349.1017B}, that is, purely radial field lines whose polarity reverses across the line of equation $\phi=\Omega t\pm \pi/2$ (Fig.~\ref{fig_setup}). The split-monopole has the advantage of depending only on radius ($\partial B/\partial\theta=0$) meaning that this solution does not exhibit the pathologies that we observed with the dipole. This configuration is also well motivated physically because it is a good approximation of the asymptotic structure of the pulsar wind, which is the main focus of this study.

As soon as the simulation begins and at every timestep, electron-positron pairs are uniformly created at the surface of the star to model the polar-cap discharge \citep{2013MNRAS.429...20T, 2014ApJ...795L..22C, 2015ApJ...815L..19P}, which was shown by \citet{2015ApJ...801L..19P} to be more efficient for inclined rotators and hence most relevant to this study. Pair creation {\em via} photon-photon annihilation away from the star is neglected here. The injected plasma is neutral (one pair of macroparticles is injected per cell and per timestep) and has a high multiplicity, $\kappa_{\star}\equiv n_{\star}/n_{\rm GJ}$, where $n_{\rm GJ}\approx \boldsymbol{\Omega}\cdot\mathbf{B}/2\pi e c$ is the fiducial Goldreich-Julian density \citet{1969ApJ...157..869G}. In our set of simulations, $\kappa_{\star}=5$, $10$, and $20$. The high plasma multiplicity allows to reach the quasi force-free regime in the magnetosphere and wind ({\em i.e.}, $\rho \mathbf{E}+\mathbf{J}\times\mathbf{B}/c=0$ with $\rho,\mathbf{J}$ the charge and current densities), except in the current sheet where reconnection operates. It is worth noting that the plasma will remain neutral everywhere in the box ($\rho\approx 0$) because both species play a symmetric role by construction of the numerical setup used here. This is most appropriate to describe an orthogonal rotator. In the more general case where $\boldsymbol{\Omega}\cdot\mathbf{B}\neq 0$, the plasma in the magnetosphere and the current sheet is charged.

The magnetization of the plasma at the surface of the star is very high,
\begin{equation}
\sigma_{\star}\equiv \frac{B^2_{\star}}{4\pi n_{\star}m_{\rm e} c^2}\gg 1,
\end{equation}
where $m_{\rm e}$ is the rest mass of the electron. In this work, $\sigma_{\star}$ varies from $250$ up to $2500$. At $r=r_{\star}$, the plasma skin-depth is smallest and is resolved by $4.4$ down to $1.4$ cells depending on the value of $\sigma_{\star}$. The highest plasma frequency $\omega^{\rm max}_{\rm pe}=c/d^{\rm min}_{\rm e}$ is well resolved in all cases, with $\omega^{-1}_{\rm pe}/\Delta t\approx 12$, and down to $4$ at the highest $\sigma_{\star}$. The shortest radiative cooling timescale must also be well resolved by the simulations, $t^{\star}_{\rm rad}\equiv 9 m_{\rm e}c/4f_{\rm rad}r^2_{\rm e}B_{\star}^2$, where $r_{\rm e}=e^2/m_{\rm e}c^2$ is the classical radius of the electron and $f_{\rm rad}$ is a numerical factor that scales the magnitude of the cooling strength. Here, $f_{\rm rad}B_{\star}^2$ is a fixed number, meaning that $t^{\star}_{\rm rad}\approx 13 \Delta t$ in all the simulations regardless of the magnetic field strength. Simulations ran until the wind reached the outer boundary, $r=150 R_{\rm LC}$, which corresponds to $\approx 24$ spin periods. At this point, the box contained about $2.75$ billions particles which represents about 160 particles per cell if spread uniformly over the box. Table~\ref{tab_sim} lists the set of simulations and the physical parameters that we explored in this study.

\begin{table}
\caption{List of numerical simulations performed in this work. The parameters are the ratio of the light-cylinder radius to the star radius, the magnetization and the multiplicity of the plasma injected at the surface of the star.}
\label{tab_sim}
\begin{tabular}{c c c c}
\hline\hline
\hspace{1.0cm}Run name\hspace{1.0cm} & $~~~R_{\rm LC}/r_{\star}~~~$ & $~~~\sigma_{\star}~~~$ & $~~~~\kappa_{\star}~~~~$ \\
\hline
{\tt R3\_S250\_K10} & 3 & 250 & 10 \\
{\tt R3\_S250\_K5} & 3 & 250 & 5 \\
{\tt R3\_S250\_K20} & 3 & 250 & 20 \\
{\tt R3\_S1000\_K10} & 3 & 1000 & 10 \\
{\tt R3\_S2500\_K10} & 3 & 2500 & 10 \\
{\tt R6\_S250\_K10} & 6 & 250 & 10 \\
{\tt R9\_S250\_K10} & 9 & 250 & 10 \\
\hline\hline                              
\end{tabular}
\end{table}

\begin{figure}
\centering
\includegraphics[width=\hsize]{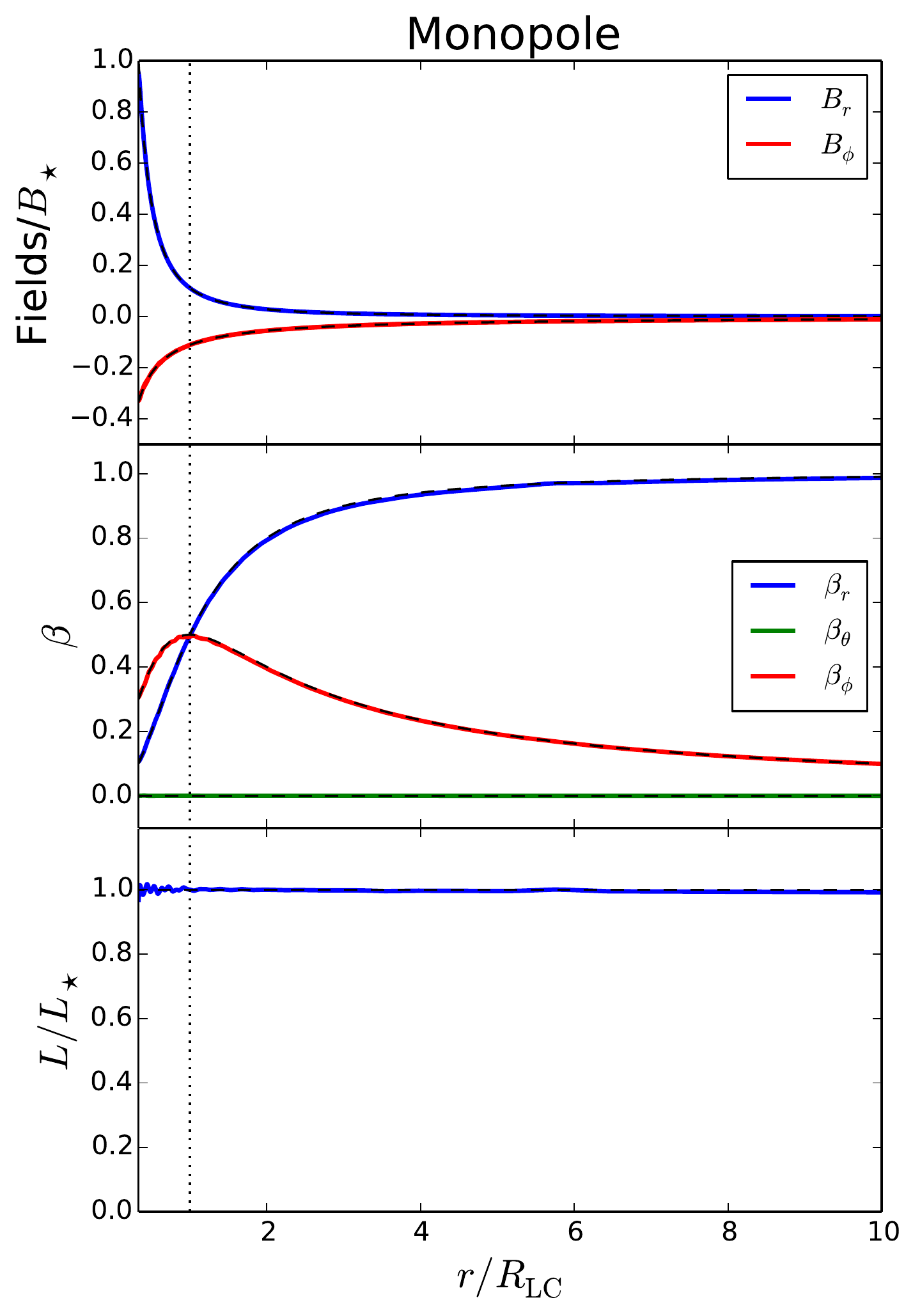}
\caption{Test simulation of a pure magnetic monopole compared with the analytical solution of \citet{1973ApJ...180L.133M} (dashed lines). From top to bottom: Non-vanishing components of the magnetic field, components of the $\boldsymbol{\beta}=\mathbf{E}\times\mathbf{B}/B^2$ drift velocity, and spindown power $L$ normalized to $L_{\star}=cB^2_{\star}r^4_{\star}/R^2_{\rm LC}$ as a function of radius.}
\label{fig_monopole}
\end{figure}

\section{Simulations of the striped wind}\label{sect_results}

\begin{figure*}
\centering
\includegraphics[width=\hsize]{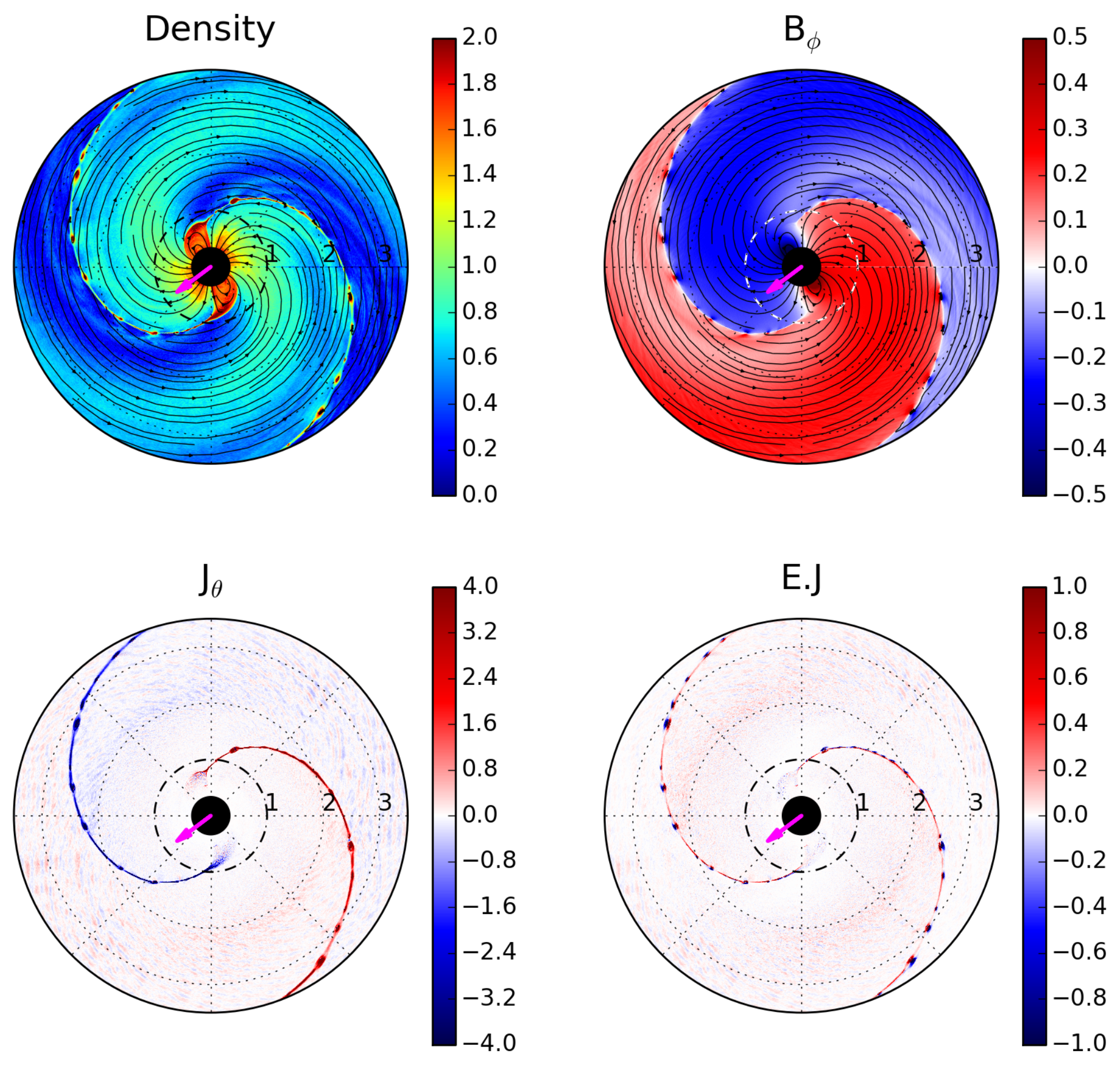}
\caption{Snapshot of the striped wind simulation zoomed in on the innermost regions ($r<3.5 R_{\rm LC}$) at time $t=16.6$ of rotation periods in run {\tt R3\_S250\_K10}. From top left to bottom right: the plasma multiplicity $n/n_{\rm GJ}$ in $\log_{10}$ scale, the toroidal magnetic field $(r/r_{\star})B_{\phi}/B_{\star}$, the poloidal current density $(r/r_{\star})^2 J_{\theta}$ normalized to the fiducial Goldreich-Julian current density $J_{\rm GJ}\approx \Omega B_{\star}/2\pi$, and the dissipation rate of the electromagnetic energy $(\mathbf{E}\cdot\mathbf{J})$ normalized by $(r_{\star}/r)^3(B_{\star} J_{\rm GJ})$. Solid lines show the magnetic field lines. They are omitted in the bottom panels for clarity. Radii are in units of the light-cylinder radius (shown here by a dashed circle). The magenta arrow indicates the direction of the north magnetic pole ({\em i.e.,} $\phi=\Omega t$).}
\label{fig_maps}
\end{figure*}

\begin{figure*}
\centering
\includegraphics[width=\hsize]{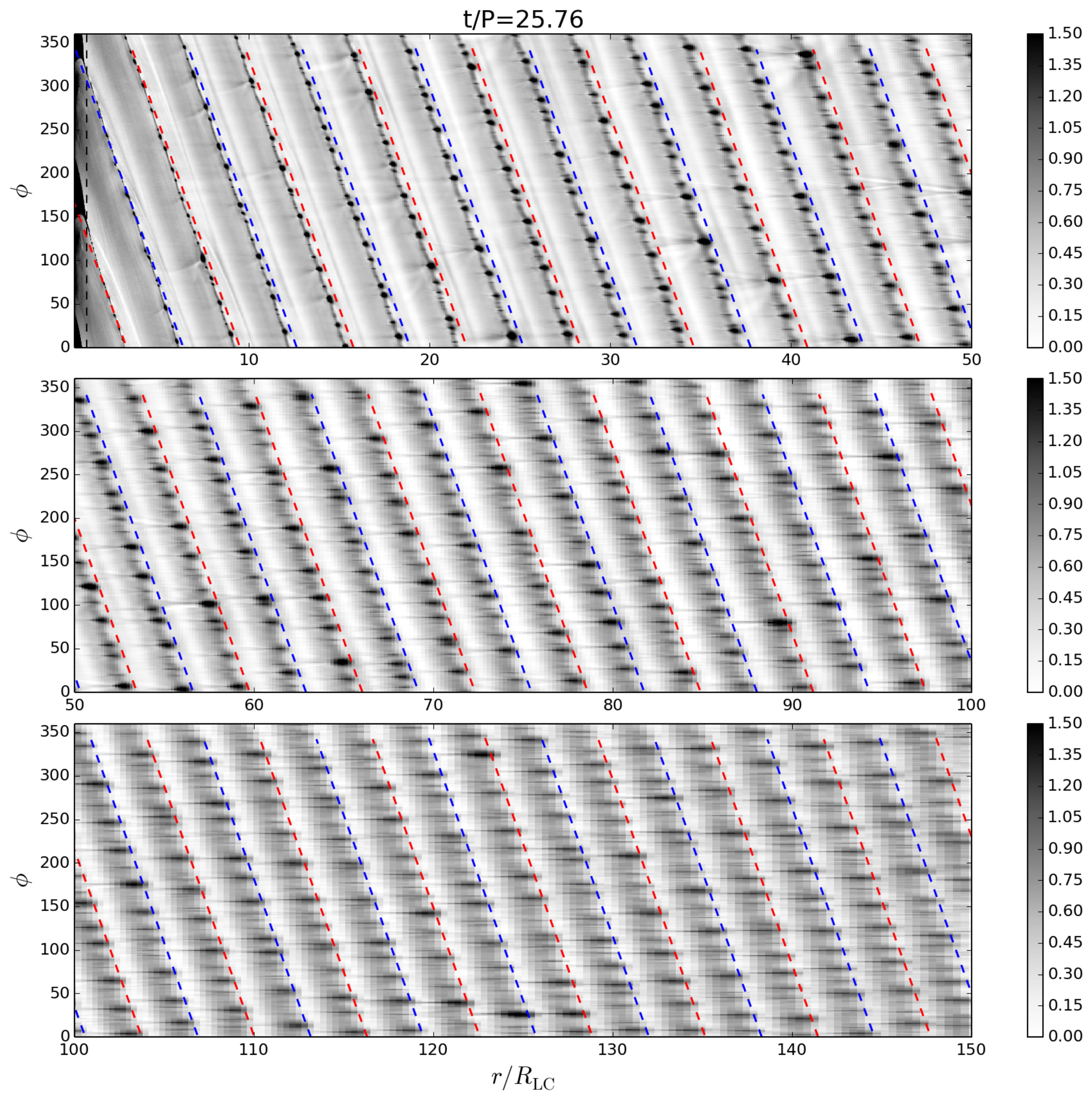}
\caption{Entire numerical domain at the end of the simulation ($t=25.76 P$) in run {\tt R3\_S250\_K10}. The grey (log-) scale shows the plasma density. Oblique dashed lines are Archimedean spirals of polar equation $r(\phi)/R_{\rm LC}=\Omega t-\phi+\pi/2$ (blue) and $r(\phi)/R_{\rm LC}=\Omega t-\phi+3\pi/2$ (red). The vertical black dashed line is at $r=R_{\rm LC}$.}
\label{fig_spirale}
\end{figure*}

\subsection{Test case: the monopole}

Before doing the full split-monopole simulations, we performed the test of a pure monopole for which we have an exact analytical solution \citep{1973ApJ...180L.133M} that we can compare with the numerical result. This solution is characterized by the remarkable feature that $E_{\theta}=B_{\phi}$ at all radii, where $B_{\phi}=B_{\rm r}\left(r/R_{\rm LC}\right)$ and $B_{\rm r}=B_{\star}\left(r_{\star}/r\right)^2$ (the other components are all zero\footnote{We note that we use the same notation for the spherical and cylindrical radii here (both written $r$) since it makes no difference in the equator.}). We recover this solution numerically with good accuracy. Figure~\ref{fig_monopole} makes a direct comparison between the analytical (in dashed lines) and the numerical solution (in solid lines). The top panel shows the radial profile of the non-zero components of the magnetic field ($B_{\rm r}$ and $B_{\phi}$). The middle panel illustrates even better the good agreement with the analytical solution. This plot compares the $\mathbf{E}\times\mathbf{B}$ drift velocity with the expected radial profiles, that is,
\begin{eqnarray}
V_{\rm r} &=& \frac{c}{1+R^2_{\rm LC}/r^2},\\
V_{\rm \theta} &=& 0,\\
V_{\phi} &=& \frac{r\Omega}{1+r^2/R^2_{\rm LC}}.
\end{eqnarray}
The bottom panel shows the radial Poynting flux going through a spherical surface as a function of radius. The expected spindown power is
\begin{equation}
L\left(r\right)=\frac{c}{4\pi} \iint{\left(\mathbf{E}\times\mathbf{B}\right)r^2\sin\theta d\theta d\phi}=\frac{cB^2_{\star}r^4_{\star}}{R^2_{\rm LC}}=L_{\star}.
\end{equation}
The numerical simulation demonstrates not only the good agreement with the analytical result, but also the low numerical dissipation of the code since no physical dissipation is expected in this particular setup. This is, of course, in contrast with the split-monopole case where reconnection kicks in and results in magnetic dissipation which is the main physical effect investigated here (see Sect.~\ref{sect_diss} below).

\subsection{Overall structure}\label{sect_struct}

We now describe the results of the striped wind simulations obtained from an initial split-monopole configuration. Figure~\ref{fig_maps} shows a snapshot of the simulation (run {\tt R3\_S250\_K10}) after $16.6$ spin periods and zoomed in on the innermost regions ($r<3.5 R_{\rm LC}$). Shortly after the onset of the simulation, the purely radial magnetic structure collapses into the more familiar structure of the pulsar magnetosphere, that is, with closed field lines in the magnetic equatorial regions ($\phi=\Omega t \pm \pi/2$) and open field lines at the magnetic poles ($\phi=\Omega t$ and $\phi=\Omega t+\pi$). The closed field line region is confined within the light cylinder (shown by the dashed circle in Fig.~\ref{fig_maps}) and is in solid rotation with the star. The striped wind begins beyond the light cylinder. It is composed of two equatorial current sheets that form at the boundary between opposite magnetic polarities (mostly in $B_{\phi}$, see top-right panel) and a cold relativistic wind of particles. The current sheets have the shape of two nested Archimedean spirals of wavelength $2\pi R_{\rm LC}$ separated from each other by $\pi R_{\rm LC}$ (most visible in Fig.~\ref{fig_spirale}). The electric current flowing in each sheet (along $\theta$ only) has opposite sign (bottom-left panel). This can easily be understood as the opposite signs of $\boldsymbol{\nabla}\times\mathbf{B}$ across each layer. The thickness of the current layers at the light cylinder, $\delta_{\rm LC}$, is approximatively the local plasma skin-depth, that is, $\delta_{\rm LC}\sim d_{\rm e}\approx 10^{-2}R_{\rm LC}$ in this run. The layer thickness is resolved by at least 6 cells at the light cylinder.

The density map in the top-left panel of Figure~\ref{fig_maps} reveals a high-density contrast between the wind zone (the inter-stripe region, with local multiplicities $\kappa\sim 1$--$10$) and the current sheets ($\kappa\sim 10$--$100$). There is also a strong density gradient between both sides of the reconnection layer, the leading part of the wind being less dense than the trailing part. The magnetic field strength follows the same trend. This resembles the asymmetric reconnection configuration as found, for example, in the Earth's magnetopause. As soon as the current sheets form, they become quickly unstable to the relativistic tearing mode \citep{1979SvA....23..460Z} which leads to the initial fragmentation of the sheet into overdense elliptical structures (plasmoids or magnetic islands below) separated by secondary current sheets (or simply referred to as X-points in the following). The fragmentation of secondary current sheets then proceeds on smaller scales and leads to the formation of a hierarchical chain of plasmoids and short current sheets {\em via} the plasmoid instability \citep{2010PhRvL.105w5002U}. The initial tearing instability was already observed in three-dimensional (3D) pulsar magnetosphere simulations \citep{2015ApJ...801L..19P, 2016MNRAS.457.2401C, 2017arXiv170704323P} but they did not capture the subsequent plasmoid instability because of the lack of sufficient scale separation in these studies.

These instabilities are also well-known in classical plane-parallel configurations to mediate efficient conversion of the Poynting flux into non-thermal particle acceleration within secondary current sheets (e.g., \citealt{2013ApJ...770..147C, 2016MNRAS.462...48S}). This point is illustrated in the bottom-right panel of Figure~\ref{fig_maps} which shows the dissipation rate of the electromagnetic energy $\mathbf{E}\cdot\mathbf{J}$. There is a net positive dissipation rate localized between plasmoids where the reconnection (non-ideal) electric field is aligned with the current. Inside magnetic islands, the wind (ideal) electric field ($\mathbf{E}= -\mathbf{V}\times\mathbf{B}/c$) which changes sign on each side of the islands dominates, resulting in a dipolar-like structure in the $\mathbf{E}\cdot\mathbf{J}$ map. There is therefore no net dissipation of the Poynting flux within the plasmoids or anywhere else in the wind outside of the secondary current sheets. We discuss the process of particle acceleration in greater detail in Sect.~\ref{sect_acc}.

Magnetic reconnection is most vigorous and time-dependent in the innermost parts of the striped wind ($r\lesssim 10 R_{\rm LC}$). Plasmoids quickly grow from the inflow of particles coming from X-points on either side and by merging with other plasmoids. Mergers between small plasmoids are frequent close to the light cylinder. At intermediate radii ($\sim 2-5 R_{\rm LC}$) merging episodes are still possible, although much less frequent, but they involve bigger structures. At larger distances, the substructures in the sheets are frozen by the adiabatic expansion. At the outer edge of the simulation box, about a dozen plasmoids remain (Fig.~\ref{fig_spirale}). The overall picture does not change with the values of initial plasma magnetization and multiplicity, and pulsar angular velocity explored here.

\subsection{Magnetic reconnection and dissipation}\label{sect_diss}

\subsubsection{Layer thickness and dissipation radius}\label{sect_delta}

\begin{figure}
\centering
\includegraphics[width=\hsize]{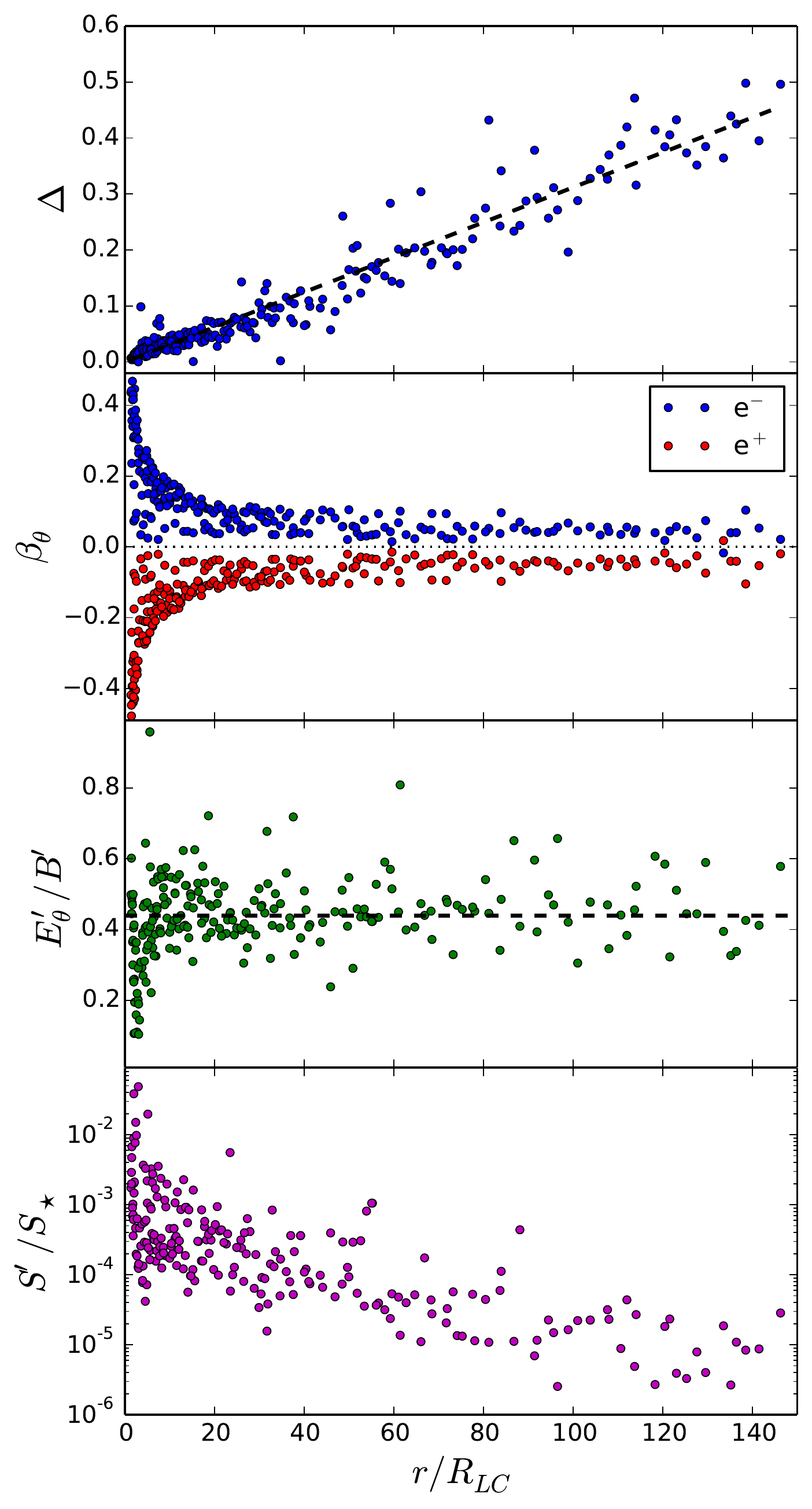}
\caption{From top to bottom: Current sheet filling factor ($\Delta=\delta/\pi R_{\rm LC}$), plasma drift velocity ($\beta_{\theta}$), dimensionless reconnection rate ($\beta^{\prime}_{\rm rec}= E^{\prime}_{\theta}/B^{\prime}$), and electromagnetic energy flux toward the sheet ($S^{\prime}=c\left|\mathbf{E}^{\prime}\times\mathbf{B^{\prime}}\right|/4\pi$ normalized by $S_{\star}=L_{\star}/4\pi R^2_{\star}$) as function of radius in run {\tt R3\_S250\_K10}. This analysis is done for each X-point (marked by a dot in these plots) along one spiral arm. The dashed line in the top panel is a linear fit to the numerical data of equation $\Delta(r)=\Delta_{\rm LC} (r/R_{\rm LC})$, where $\Delta_{\rm LC}\approx 0.003$. The horizontal dashed line gives the average dimensionless reconnection rate $\beta^{\prime}_{\rm rec}\approx 0.44$.}
\label{fig_reconnection}
\end{figure}

Figure~\ref{fig_spirale} clearly shows the broadening of the current sheets with radius. To quantify this, we measure the layer thickness $\delta$ at each X-point that we identify as a local density minimum along both spiral arms. We perform a Gaussian fit of the density profile along the direction perpendicular to the layer. We define the layer thickness as the full width at half maximum of the best-fit solution, $\delta={\rm FWHM}$. Following \citet{2001ApJ...547..437L}, we define the fraction of the striped wind wavelength filled by the current layer, $\Delta=\delta/\pi R_{\rm LC}$. We find that $\Delta$ increases linearly with radius (see top panel in Fig.~\ref{fig_reconnection}). A linear fit gives
\begin{equation}
\Delta\left(r\right)\approx \Delta_{\rm LC} \left(\frac{r}{R_{\rm LC}}\right),
\label{eq_diss}
\end{equation}
where $\Delta_{\rm LC}=\delta_{\rm LC}/\pi R_{\rm LC}\approx 3.1\times 10^{-3}$ in this particular run. We find this correlation in all runs with very similar values of $\Delta_{\rm LC}$ (see Table~\ref{tab_delta}). This result implies that by extrapolation the sheet would fill the entire wind region at $r_{\rm diss}/R_{\rm LC}=\Delta^{-1}_{\rm LC}$ ({\em i.e.}, where $\Delta=1$), which would in turn lead to complete dissipation of the stripes. In this run, this would happen beyond $\gtrsim 300 R_{\rm LC}$ which is unfortunately outside the numerical box.

\begin{table}
\caption{All runs present a linear expansion of the current sheet filling factor with radius, $\Delta=\Delta_{\rm LC}\left(r/R_{\rm LC}\right)$, where the values of $\Delta_{\rm LC}$ measured in each simulation are given below.}
\label{tab_delta}
\begin{tabular}{c c}
\hline\hline
\hspace{1.0cm}Run name\hspace{1.0cm} & $~~~\Delta_{\rm LC}~~~$ \\
\hline
{\tt R3\_S250\_K10} & $3.1\times10^{-3}$ \\
{\tt R3\_S250\_K5} &  $2.8\times10^{-3}$ \\
{\tt R3\_S250\_K20} & $2.7\times10^{-3}$ \\
{\tt R3\_S1000\_K10} & $3.5\times10^{-3}$ \\
{\tt R3\_S2500\_K10} & $3.4\times10^{-3}$ \\
{\tt R6\_S250\_K10} & $3.8\times10^{-3}$ \\
{\tt R9\_S250\_K10} & $3.6\times10^{-3}$ \\
\hline\hline                              
\end{tabular}
\end{table}

The linear expansion of the sheet is consistent with the striped wind model of \citet{1990ApJ...349..538C} and \citet{1994ApJ...431..397M} which we briefly repeat here for completeness. Amp\`ere's law across the sheet for $r\gg R_{\rm LC}$ gives
\begin{equation}
2B_{\phi}=\frac{4\pi}{c}J_{\theta}\delta=4\pi n e\left(\beta_{\theta}^{+}-\beta_{\theta}^-\right)\delta,
\label{eq_ampere}
\end{equation}
where $\beta_{\theta}^{+}$ and $\beta_{\theta}^{-}$ are, respectively, the bulk drift velocity of the positrons and the electrons carrying the current\footnote{We note that this expression assumes a neutral plasma, $n_{\rm e^+}\approx n_{e^-}$, which is the case in this setup (see Sect.~\ref{sect_setup}).}. Since $B_{\phi}\propto 1/r$ on the left-hand side and the plasma density $n\propto 1/r^2$ on the right-hand side, the product $\beta_{\theta}\delta$ should linearly increase with radius. Our analysis shown in Figure~\ref{fig_reconnection} (middle panel) reveals that the plasma drift velocity is approximatively constant for $r\gg R_{\rm LC}$, $\beta_{\theta}^{+}-\beta_{\theta}^{-}\approx 0.1$, which explains why $\Delta\propto r$. We notice that both species are perfectly counter-streaming in the sheet at all radii. The drift velocity is higher close to the light cylinder ($\beta_{\theta}^{+}-\beta_{\theta}^{-}\approx 0.9$ at $r\approx R_{\rm LC}$) because $B_{\rm r}$ cannot be neglected there and participates to the reconnecting field. This result also implies that the sheet always has enough charge to conduct the current required to maintain the discontinuity of the field (since $\beta_{\theta}\ll 1$) as long as there is enough charge to begin with, as suggested by \citet{2012SSRv..173..341A} but in contradiction with \citet{1975Ap&SS..32..375U, 1996MNRAS.279.1168M}. The same conclusion holds with the other values of $\sigma_{\star}$, $\kappa_{\star}$, and $R_{\rm LC}$ explored here.

\begin{figure}
\centering
\includegraphics[width=\hsize]{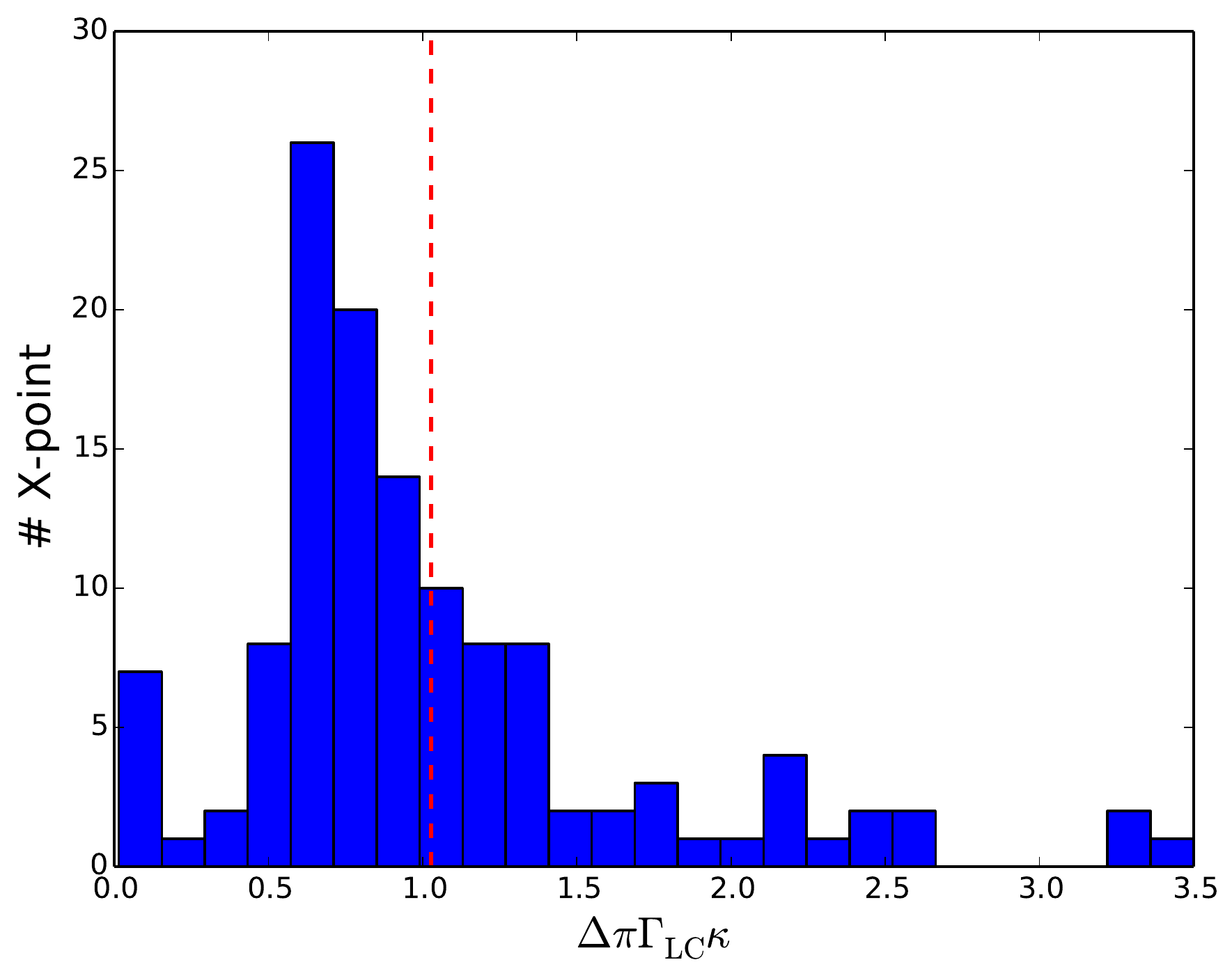}
\caption{Distribution of the product $\Delta\pi\Gamma_{\rm LC}\kappa$ computed at each X-point along the striped wind (run {\tt R3\_S250\_K10}). The distribution peaks at $\Delta\pi\Gamma_{\rm LC}\kappa\approx 0.7$ with a mean value of order unity (vertical red dashed line) as expected from Eq.~(\ref{eq_kappa}).}
\label{fig_delta}
\end{figure}

An important consequence of Eq.~(\ref{eq_diss}) is that the distance of complete dissipation is solely given by the thickness of the sheet at the light cylinder. Then, the relevant question is how does $\delta_{\rm LC}$ depend on physical parameters. Using Eq.~(\ref{eq_ampere}), and assuming that $\beta_{\theta}^{+}-\beta_{\theta}^-\approx 1$ at the light cylinder (Fig.~\ref{fig_reconnection}), gives
\begin{equation}
\delta_{\rm LC}=\frac{B_{\rm LC}}{2\pi \Gamma_{\rm LC} n^{\prime}_{\rm LC} e}\approx \frac{R_{\rm LC}}{\Gamma_{\rm LC}\kappa_{\rm LC}},
\label{eq_delta}
\end{equation}
where $\Gamma_{\rm LC}$ and $n^{\prime}_{\rm LC}=n_{\rm LC}/\Gamma_{\rm LC}$ are the bulk Lorentz factor and the plasma proper density at the light cylinder, respectively. The plasma multiplicity parameter is defined as $\kappa_{\rm LC}=n^{\prime}_{\rm LC}/n_{\rm GJ}$. Then, Eq.~(\ref{eq_diss}) simply becomes
\begin{equation}
\Delta\left(r\right)\approx\frac{1}{\pi\Gamma_{\rm LC}\kappa_{\rm LC}}\left(\frac{r}{R_{\rm LC}}\right)=\frac{1}{\pi\Gamma_{\rm LC}\kappa},
\label{eq_kappa}
\end{equation}
where the local plasma multiplicity is $\kappa=n^{\prime}/n_{\rm GJ}=\kappa_{\rm LC}\left(R_{\rm LC}/r\right)$. This relation holds quite well at all radii in all the simulations within the range of the parameters explored here (see Fig.~\ref{fig_delta}). The plasma multiplicity and bulk Lorentz factor at the light cylinder are about the same in all runs, $\kappa_{\rm LC}\sim 100$ and $\Gamma_{\rm LC}\approx 1$ (see Sect.~\ref{sect_rate}), which explains why the layer thickness is very similar in all cases (even if $\kappa_{\star}$ and $\sigma_{\star}$ change). We reach the important conclusion that the striped wind would come to a complete dissipation at (where $\Delta=1$)
\begin{equation}
\frac{r_{\rm diss}}{R_{\rm LC}}=\pi \Gamma_{\rm LC} \kappa_{\rm LC},
\label{eq_rdiss}
\end{equation}
which is identical to Eq.~(11) in \citet{2001ApJ...547..437L}.

\begin{figure}
\centering
\includegraphics[width=\hsize]{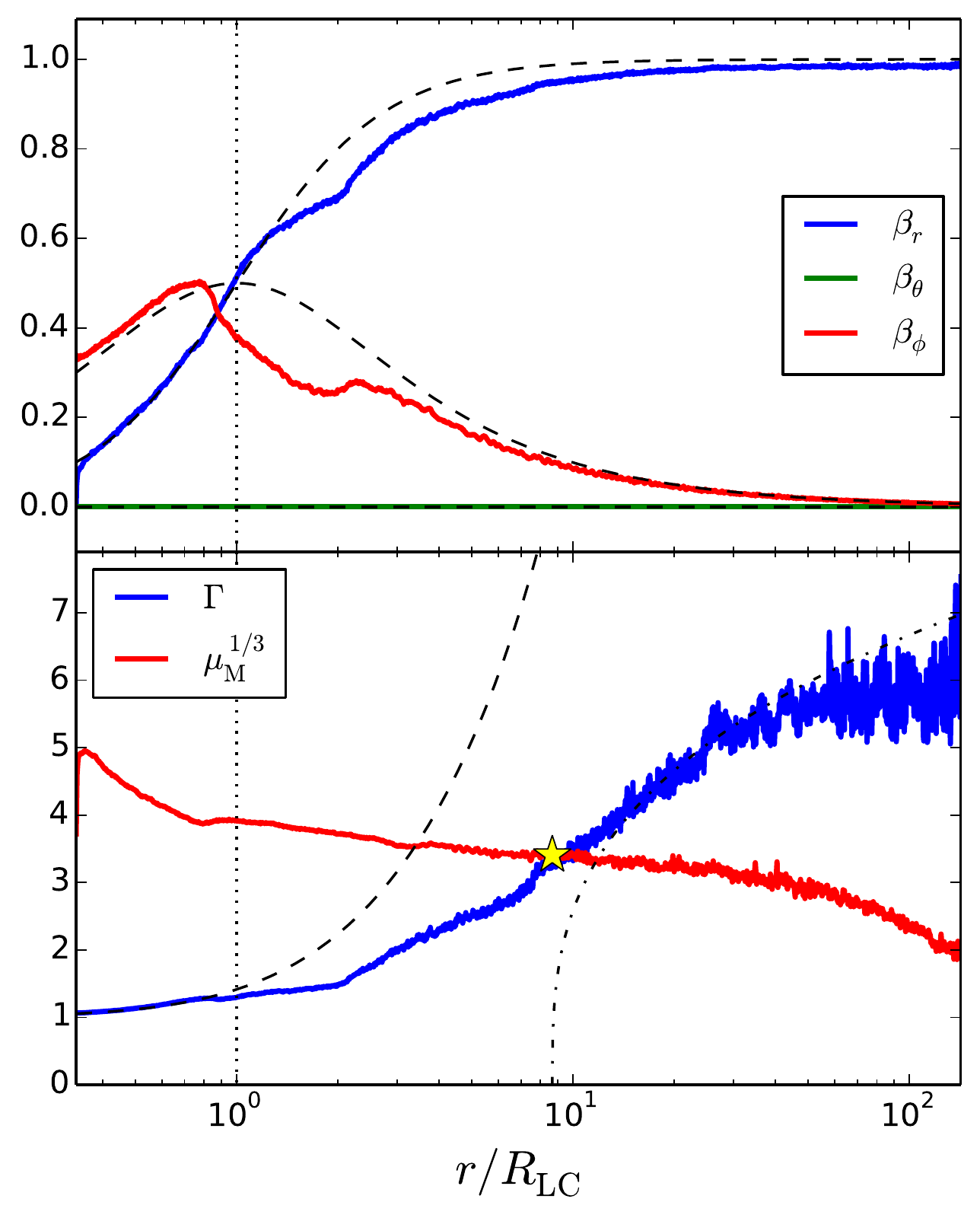}
\caption{Top: Radial profile of the plasma bulk velocity components $\mathbf{V}/c$ of the striped wind averaged in $\phi$ (solid lines) and the analytical $\mathbf{E}\times\mathbf{B}$ drift velocity of a pure monopole (dashed lines) at $t=25.76 P$ in run {\tt R3\_S250\_K10}. Bottom: Corresponding $\phi$-averaged bulk Lorentz factor of the striped wind (blue line). The wind first accelerates approximatively as $(1+r^2/R^2_{\rm LC})^{1/2}$ (dashed line) and then grows as $\sigma^{1/3}_{\star}\ln^{1/3}(r/r_{\rm FMS})$ (dot-dashed line) passed the fast point (yellow star) where the $\mu^{1/3}_{\rm M}$ (red line) and the $\Gamma$ curves intersect.}
\label{fig_wind}
\end{figure}

\subsubsection{Reconnection rate and bulk velocities}\label{sect_rate}

Another key quantity to measure along the sheet is the dimensionless reconnection rate. It is usually defined as the ratio of the reconnection electric field induced at X-points and the upstream reconnecting magnetic field ({\em i.e.}, just outside the layer). Equivalently, it is given by the speed at which fresh plasma and field enter and then leave an X-point, that is, as the ratio of the inflow to outflow plasma bulk velocities $V_{\rm in}/V_{\rm out}$. It is straightforward to measure this quantity in a 2D parallel configuration where the sheet has no bulk motion. In this context, the analysis is more complex because the relativistic bulk motion of the wind must be subtracted to extract the mildly relativistic motion induced by reconnection in the vicinity of each X-point. It is thus appropriate to measure the reconnection rate in the wind comoving frame \citep{1996A&A...311..172L, 2014ApJ...780....3U}. Another difficulty is that the wind accelerates with radius and is not uniform along $\phi$. Figure~\ref{fig_wind} shows the radial profile of the plasma bulk velocity averaged in $\phi$. These profiles are consistent with the $\mathbf{E}\times\mathbf{B}$ drift velocity of a pure monopolar relativistic wind (see Fig.~\ref{fig_monopole}), but with some deviations close to the light-cylinder due to reconnection. The bulk Lorentz factor grows with radius, approximatively as
\begin{equation}
\Gamma\approx\sqrt{1+\frac{r^2}{R^2_{\rm LC}}},
\end{equation}
close to the light cylinder \citep{1973ApJ...180L.133M, 1977MNRAS.180..125B, 2002ApJ...566..336C}, but quickly saturates to $\Gamma\approx 5$-$6$ at large radii ($r\gtrsim 20$). The acceleration of the wind slows down after crossing the fast magnetosonic point located at $r=r_{\rm FMS}\approx 8.7 R_{\rm LC}$ and $\Gamma_{\rm FMS}\approx 3.4$ for $\sigma_{\star}=250$ (indicated by the yellow star in Fig.~\ref{fig_wind}), that is, where the wind velocity exceeds the Alfv\'en speed. This is given by the condition $\Gamma=\mu_{\rm M}^{1/3}$, where 
\begin{equation}
\mu_{\rm M}=\frac{B^2}{4\pi n m_{\rm e} c^2}=\Gamma\sigma,
\end{equation}
\citep{1969ApJ...158..727M, 2009ASSL..357..421K}. The plasma inertia becomes important beyond this point and the wind Lorentz factor increases only logarithmically with radius, approximatively as
\begin{equation}
\Gamma\approx \sigma^{1/3}_{\star}\ln^{1/3}\left(\frac{r}{r_{\rm FMS}}\right),
\end{equation}
in agreement with previous studies \citep{1994PASJ...46..123T, 1998MNRAS.299..341B, 2009ApJ...699.1789T, 2009ApJ...698.1570L}. We see the same behavior in the other runs but, as expected, the distance of the fast point to the star increases with increasing magnetization: $r_{\rm FMS}\approx 33 R_{\rm LC}$ for $\sigma_{\star}=1000$ and $r_{\rm FMS}\approx 77 R_{\rm LC}$ for $\sigma_{\star}=2500$.

We use the $\phi$-averaged bulk velocity profile shown in Figure~\ref{fig_wind} to perform a Lorentz boost on each point of the striped wind, and neglect the effects of acceleration of the wind. Fields are changed according to the usual Lorentz transformation formulae, that is,
\begin{eqnarray}
\mathbf{E}^{\prime} &=& \Gamma\left(\mathbf{E}+\boldsymbol{\beta}\times\mathbf{B}\right)-\frac{\left(\Gamma-1\right)}{\beta^2}\left(\mathbf{E}\cdot\boldsymbol{\beta}\right)\boldsymbol{\beta},\\
\mathbf{B}^{\prime} &=& \Gamma\left(\mathbf{B}-\boldsymbol{\beta}\times\mathbf{E}\right)-\frac{\left(\Gamma-1\right)}{\beta^2}\left(\mathbf{B}\cdot\boldsymbol{\beta}\right)\boldsymbol{\beta},
\end{eqnarray}
where primed quantities are defined in the comoving frame. Far from the sheets, the electric field vanishes $\mathbf{E}^{\prime}\approx 0$ and the magnetic field shrinks as $\mathbf{B}^{\prime}\approx\mathbf{B}/\Gamma$, as expected if $\mathbf{E}\cdot\mathbf{B}=0$. We compute the dimensionless reconnection rate, $\beta^{\prime}_{\rm rec}$, as the ratio of the electric field induced inside and around each X-point by reconnection, $E^{\prime}_{\theta}$, with the upstream magnetic field strength $B^{\prime}$. We obtain a dimensionless reconnection rate around $0.4$ throughout the wind (see bottom panel in Fig.~\ref{fig_reconnection}). These values are high compared with previous local plane parallel simulations of reconnection (with a rate of order 0.1-0.2). A possible explanation is that reconnection is driven by large-scale plasma motions in the wind rather than being spontaneous as is almost always the case in local simulations. Intense synchrotron losses may also be at the origin of this enhancement as anticipated by \citet{2011PhPl...18d2105U} and as observed in \citet{2013ApJ...770..147C}. Figure~\ref{fig_reconnection} (bottom panel) shows the module of the Poynting vector defined in the comoving frame, $S^{\prime}=c\left|\mathbf{E}^{\prime} \times \mathbf{B}^{\prime} \right|/4\pi$, normalized by $S_{\star}=L_{\star}/4\pi R^2_{\star}$. This quantity measures the amount of electromagnetic energy flux flowing towards X-points along the striped wind. Even though the reconnection rate remains constant, the available energy flux sharply decreases with radius because $S^{\prime}\approx c\beta^{\prime}_{\rm rec}B^2_{\phi}/\Gamma^2\propto 1/r^{2}$ if $\Gamma$ is constant.

\begin{figure}
\centering
\includegraphics[width=\hsize]{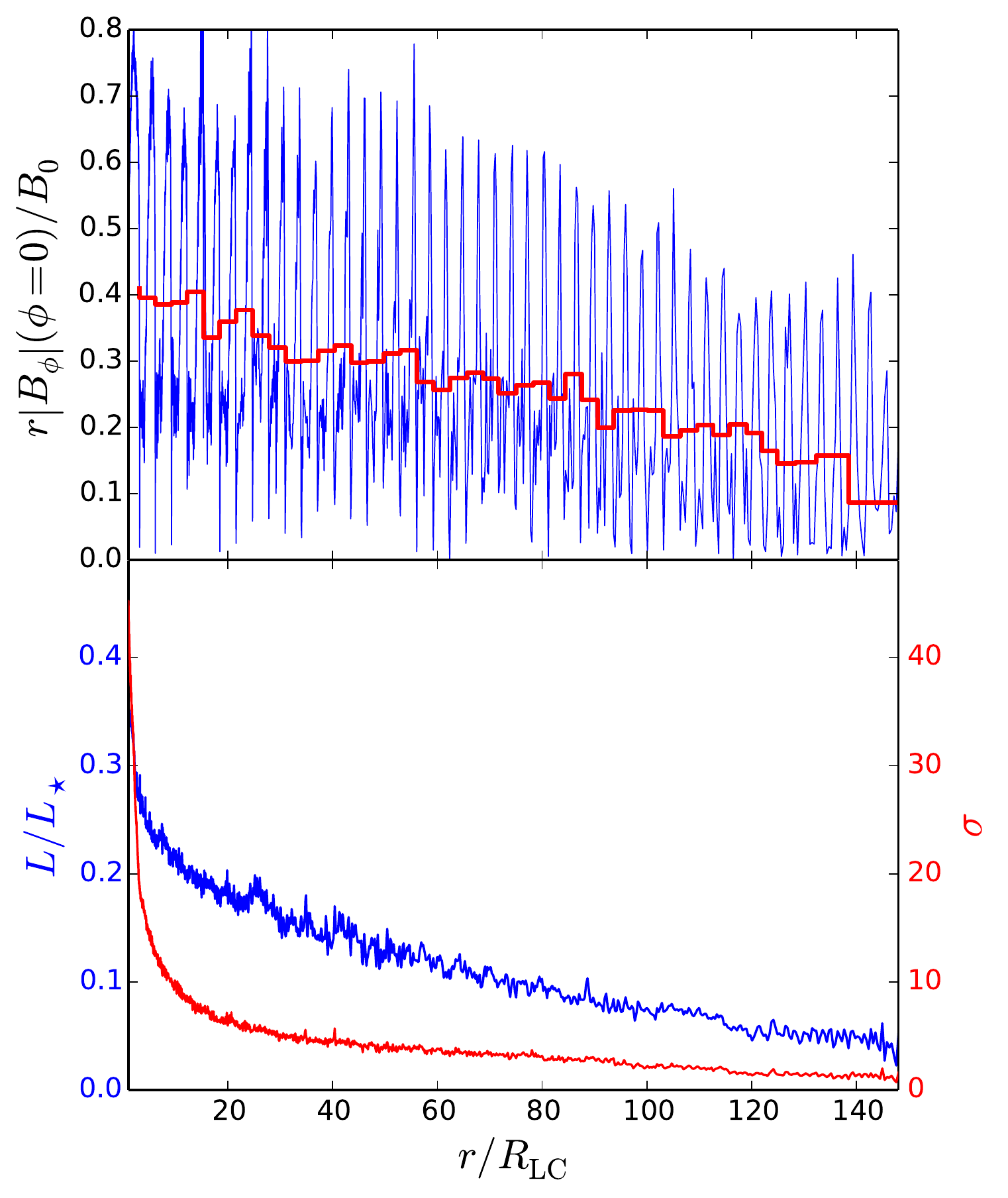}
\caption{Top: Radial profile of the toroidal magnetic field strength $r\left|B_{\phi}\right|$ (blue line) and its striped-average (red line) for $\phi=0$ at $t=25.76 P$ in run {\tt R3\_S250\_K10}. Bottom: Radial profile of the Poynting flux ($L/L_{\star}$, blue line) and the plasma magnetization parameter ($\sigma=B^2/4\pi\Gamma n m_{\rm e}c^2$, red line).}
\label{fig_dissipation}
\end{figure}

Relativistic magnetic reconnection leads to efficient magnetic dissipation. Figure~\ref{fig_dissipation} shows the decay with radius of the stripe-averaged toroidal magnetic field strength consumed by reconnection in the sheet and converted into heat and particle acceleration in run {\tt R3\_S250\_K10}. This translates into a monotonic decrease of the Poynting flux with radius, from $L\approx 0.4 L_{\star}$ at the light cylinder down to $L\approx 0.05 L_{\star}$ at $r=150 R_{\rm LC}$ ({\em i.e.}, a drop of almost 90\%), without any sign of saturation. The plasma magnetisation drops accordingly with increasing radius down to $\sigma$ of order unity at $r=150~R_{\rm LC}$, due to both increasing wind Lorentz factor and decreasing magnetic field strength. This is also consistent with the increase of the sheet filling factor with radius. We notice a more efficient dissipation close to the light cylinder where the sheet is thinner and where the tearing instability is most active. At large distances, reconnection proceeds in a smoother, more laminar way which results in an approximate constant dissipation rate.

\subsection{Particle acceleration and radiative signatures}\label{sect_acc}

\begin{figure}
\centering
\includegraphics[width=\hsize]{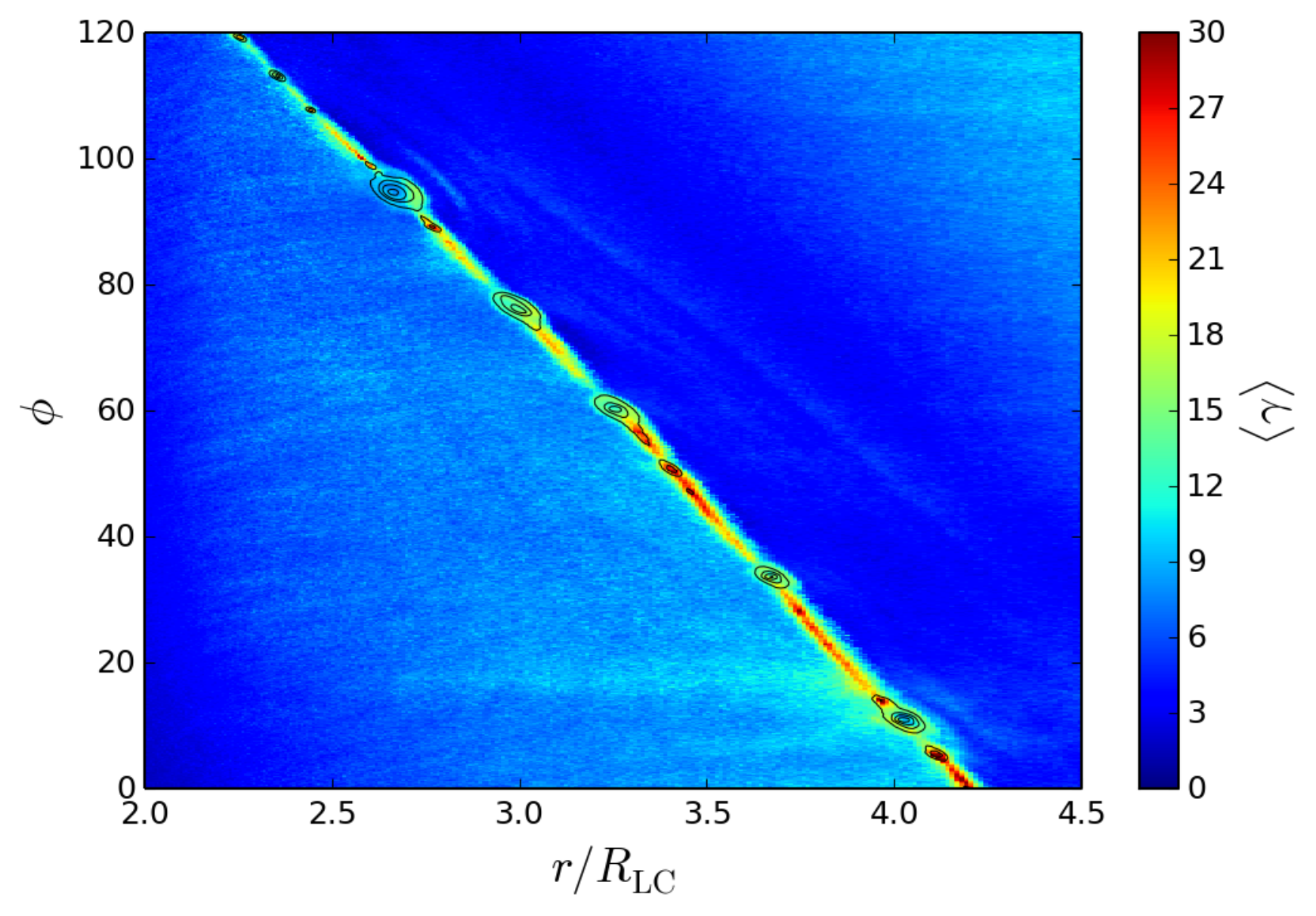}
\includegraphics[width=\hsize]{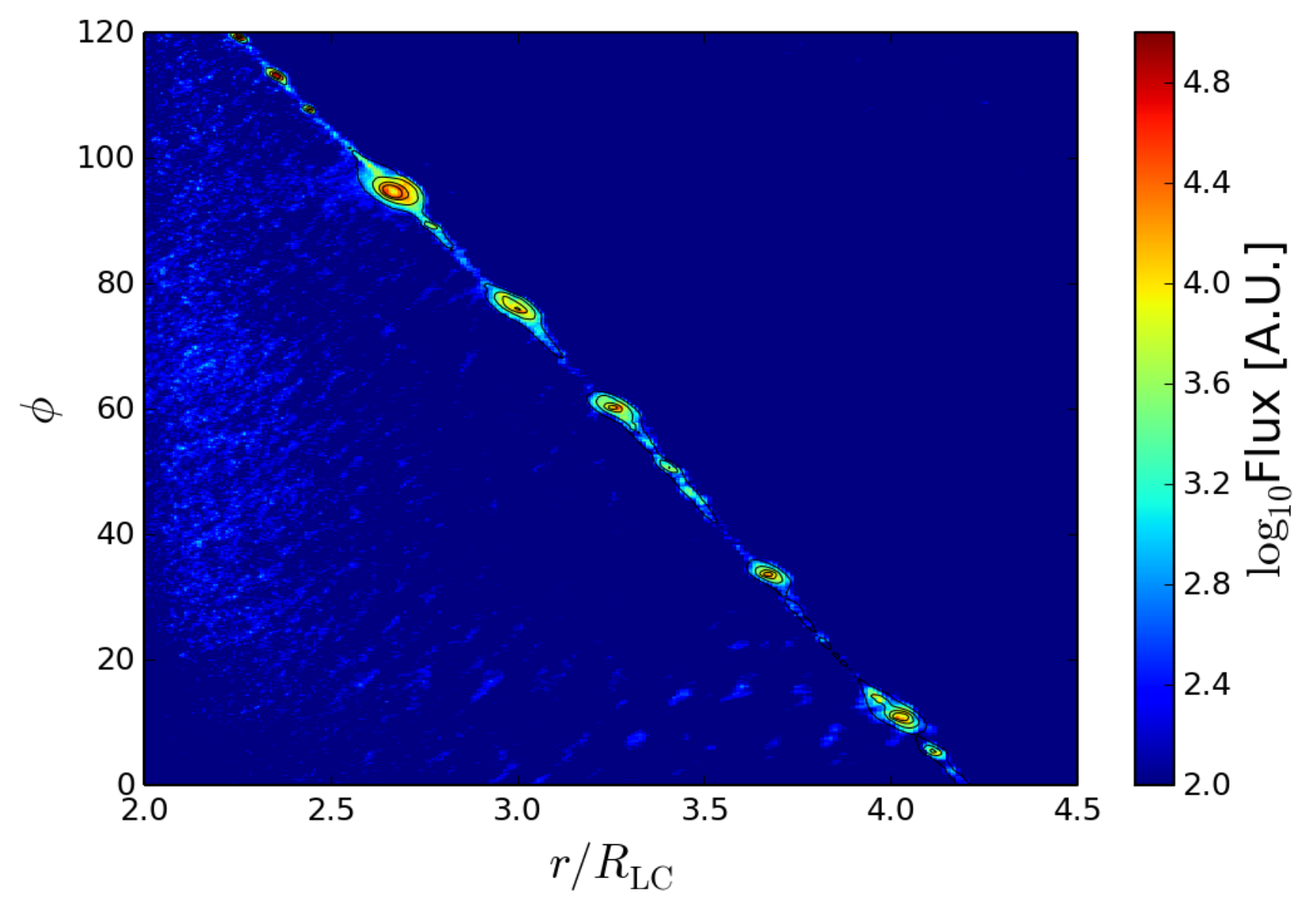}
\caption{Spatial distribution of the particle mean Lorentz factor (top) and synchrotron flux (bottom, in arbitrary units) zoomed in on the innermost regions of the striped wind in run {\tt R3\_S250\_K10}. The plasma density isocontours (black lines) show the location of plasmoids.}
\label{fig_sheet}
\end{figure}

\begin{figure}
\centering
\includegraphics[width=\hsize]{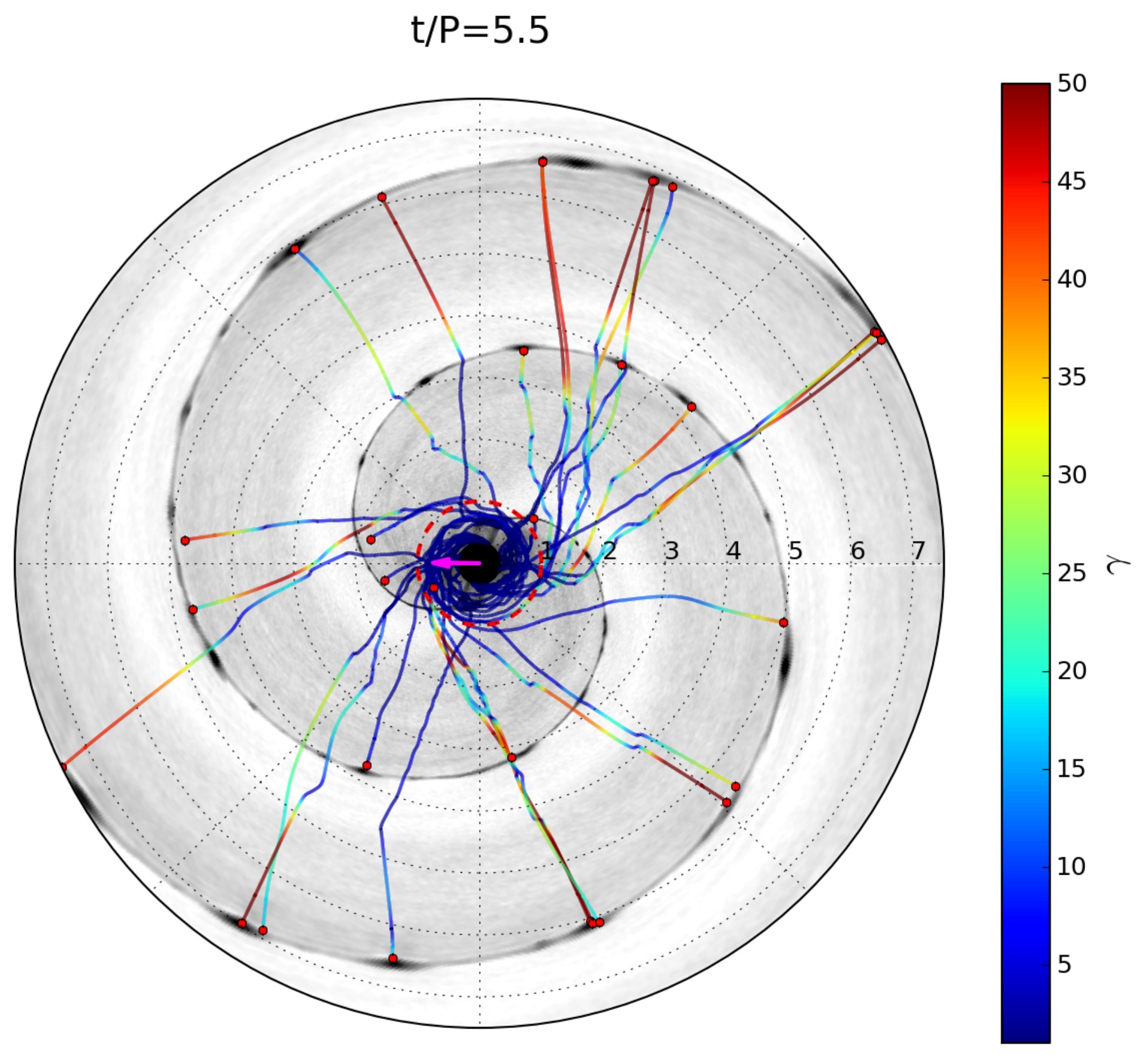}
\caption{Trajectories of 30 simulation particles whose Lorentz factor (color-coded) exceeded $\gamma=30$ at least once over their history in run {\tt R3\_S250\_K10}. Plasma density at $t=5.5P$ is shown in gray.}
\label{fig_orbit}
\end{figure}

The magnetic energy dissipated {\em via} reconnection is channeled into particle kinetic energy. As was already noticed in the map of $\mathbf{E}\cdot\mathbf{J}$ in Figure~\ref{fig_maps}, dissipation of the Poynting flux happens primarily between plasmoids, that is, within secondary current layers where the reconnection electric field accelerates particles. This picture is consistent with local plane parallel PIC simulations \citep{2001ApJ...562L..63Z, 2012ApJ...754L..33C, 2014ApJ...783L..21S, 2015ApJ...815..101N} which identified X-points as the main particle accelerating regions.

Figure~\ref{fig_sheet} (top) shows the spatial distribution of the mean particle Lorentz factor zoomed in on the innermost regions of the striped wind. This Figure clearly demonstrates that close to the light cylinder high-energy particles are concentrated within secondary layers only and not in plasmoids, in contrast with previous reconnection studies. This important difference is attributed to the strong radiative losses present in this study. Particles are first accelerated at X-points where the effective perpendicular magnetic field, given by
\begin{equation}
\tilde{B}_{\perp}=\sqrt{\left(\mathbf{E}+\boldsymbol{\beta}\times\mathbf{B}\right)^2-\left(\boldsymbol{\beta}\cdot\mathbf{E}\right)^2},
\end{equation}
is small and, hence, radiative losses are small ($\boldsymbol{\beta}$ is the particle 3-velocity divided by c). But as soon as the particles reach a plasmoid, they feel an abrupt increase of $\tilde{B}_{\perp}$ which results in catastrophic radiative energy losses $\propto \gamma^2 \tilde{B}^2_{\perp}$ (where $\gamma=(1-\beta^2)^{-1/2}$ is the particle Lorentz factor) and an intense emission of energetic synchrotron radiation \citep{2013ApJ...770..147C, 2014ApJ...782..104C}. In other words, plasmoids act as beam dumps and are the main emitting regions of the striped wind. This is well visible in Figure~\ref{fig_sheet} which shows an almost perfect spatial anti-correlation between the synchrotron flux and the mean Lorentz factor maps.

Figure~\ref{fig_orbit} presents the trajectories of $30$ simulation particles which were accelerated at least once over their history. A closer look at individual trajectories reveal multiple episodes of acceleration and cooling which depend on the evolution of the particles within the layer, and on the dynamics of plasmoids and mergers. Kinks in the trajectories are almost always associated with the capture of the particles by a plasmoid and by rapid radiative losses. We note that this connection cannot be directly observed in Figure~\ref{fig_orbit} because the particle trajectories are over-plotted onto the density map shown at a given time ($5.5$ rotation periods).

\begin{figure}
\centering
\includegraphics[width=\hsize]{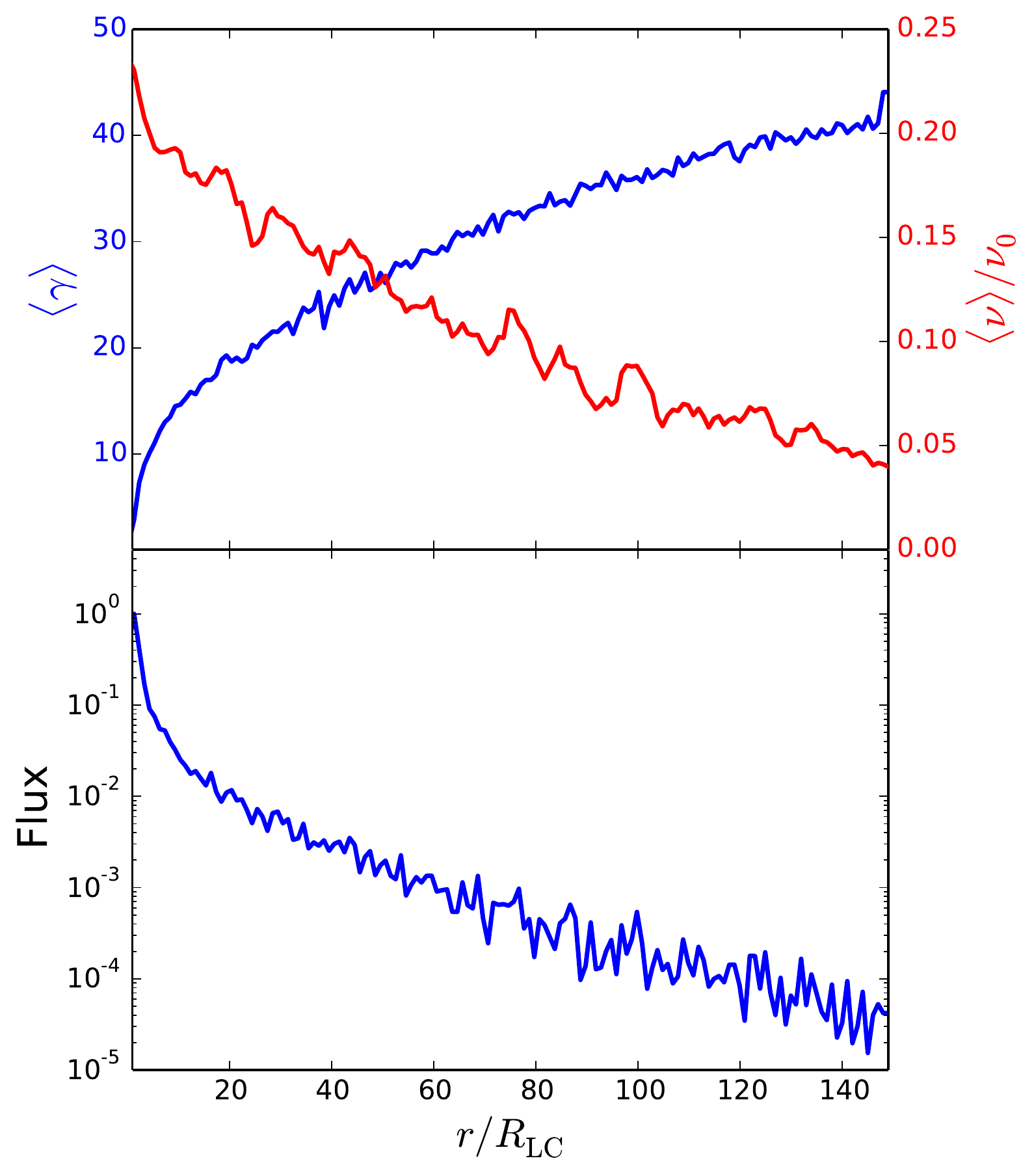}
\caption{Top: Mean particle Lorentz factor ($\langle\gamma\rangle$, blue line) and mean synchrotron photon energy ($\langle\nu\rangle/\nu_0$, where $\nu_0=3eB_{\star}/4\pi m_{\rm e}c$, red line) as functions of radius in run {\tt R3\_S250\_K10}. Bottom: Radial profile of the synchrotron flux normalized by its maximum value.}
\label{fig_energy}
\end{figure}

\begin{figure}
\centering
\includegraphics[width=\hsize]{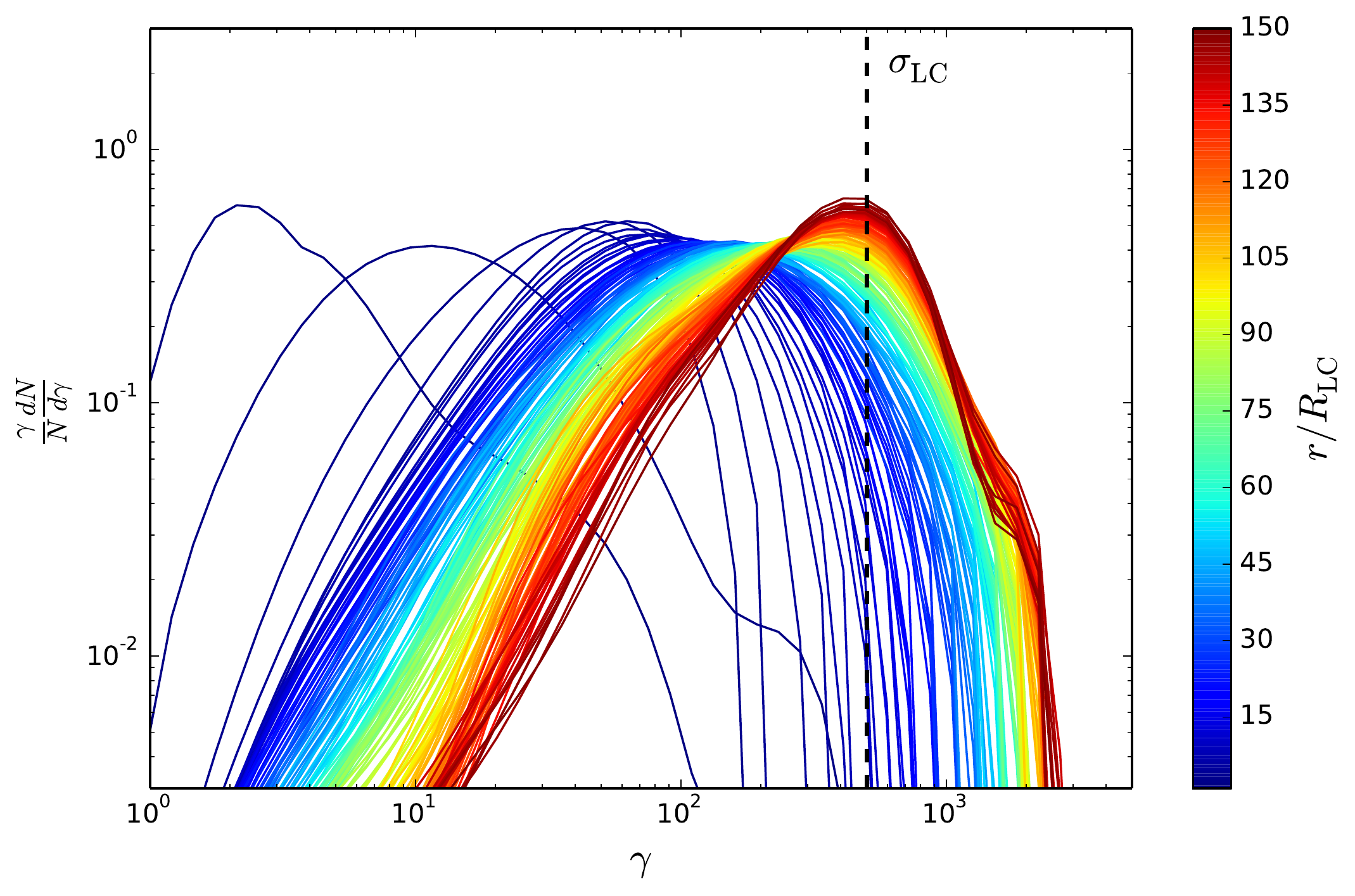}
\includegraphics[width=\hsize]{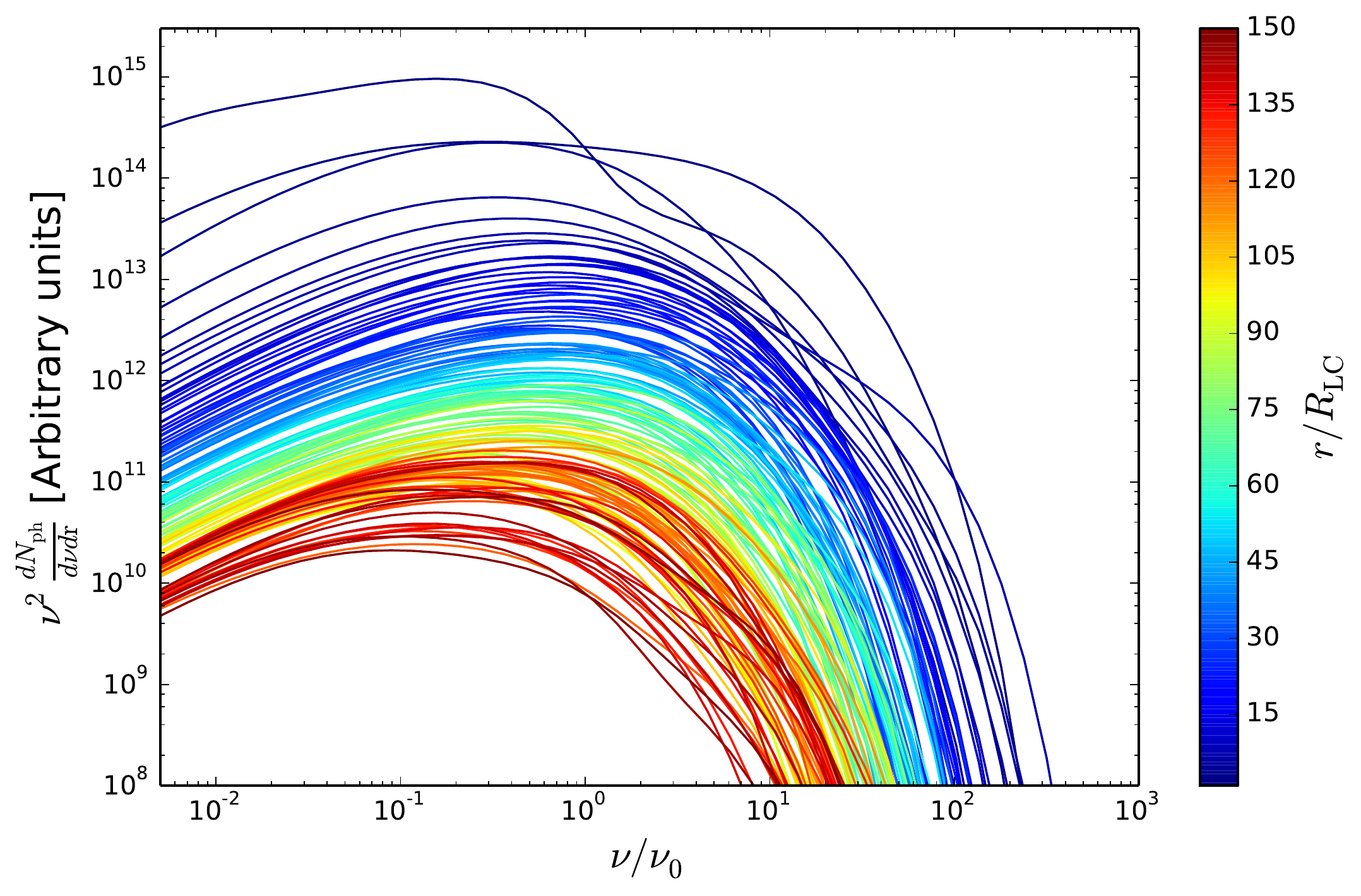}
\caption{Particle energy spectra ($\gamma dN/d\gamma$, top) and spectral energy distributions ($\nu F\nu$, bottom) averaged in $\phi$ as a function of radius (color-coded) for $\sigma_{\star}=2500$ (run {\tt R3\_S2500\_K10}). The vertical dashed line in the top panel shows $\gamma=\sigma_{\rm LC}\approx 500$.}
\label{fig_spectra}
\end{figure}

\begin{figure}
\centering
\includegraphics[width=\hsize]{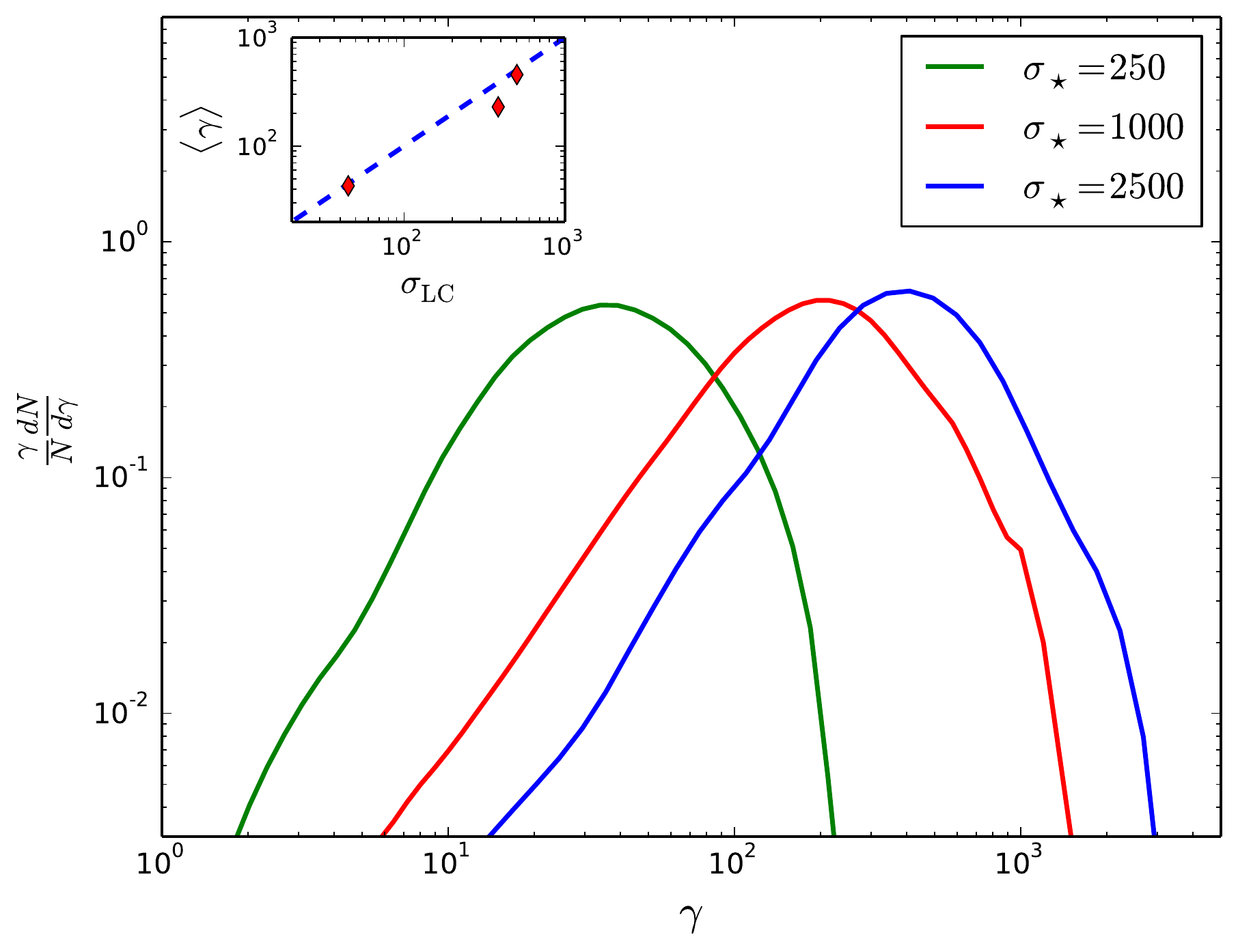}
\caption{Particle spectra at the outer edge of the simulation box $r=150 R_{\rm LC}$ for $\sigma_{\star}=250$ (green), $1000$ (red), and $2500$ (blue). The inset plot shows the mean particle Lorentz factor $\langle\gamma\rangle$ as function of $\sigma_{\rm LC}$. The oblique dashed line is $\langle\gamma\rangle=\sigma_{\rm LC}$.}
\label{fig_spectra_sigma}
\end{figure}

Further away ($r\gg R_{\rm LC}$), particles continue to accelerate because magnetic reconnection is still active and synchrotron radiative losses decrease with radius (Fig.~\ref{fig_energy}, top). Figure~\ref{fig_spectra} (top panel) shows the $\phi$-averaged particle spectra $\gamma dN/d\gamma$ as a function of radius for $\sigma_{\star}=2500$ (run {\tt R3\_S2500\_K10}). At small radii, the particle spectra are dominated by a narrow component at low energy ($\gamma\approx$ a few) which corresponds to the mild energization that experienced the particles at the surface of the star after their injection. Then, the spectrum develops a steep power-law tail that hardens with increasing radius. At large radii ($r\gtrsim 100 R_{\rm LC}$), the spectrum converges towards a narrow distribution centred around $\gamma\approx 500$, which coincides with the value of the magnetization parameter at the light-cylinder radius $\sigma_{\rm LC}\equiv \sigma(R_{\rm LC})$ \citep{2014ApJ...785L..33P, 2015MNRAS.448..606C, 2016MNRAS.457.2401C, 2017arXiv170704323P}, regardless of the strength of radiative cooling \citep{2004PhRvL..92r1101K, 2012ApJ...746..148C, 2016ApJ...833..155K}. This correlation is confirmed by the other simulations where $\langle\gamma\rangle\approx \sigma_{\rm LC}$ (see Fig.~\ref{fig_spectra_sigma}). This relation can be recast as \citep{2015MNRAS.448..606C}
\begin{equation}
\langle\gamma\rangle\approx \frac{\phi_{\rm pc}}{4\Gamma_{\rm LC}\kappa_{\rm LC}},
\end{equation}
where $\phi_{\rm pc}=e\Phi_{\rm pc}/m_{\rm e} c^2$ and $\Phi_{\rm pc}=\mu\Omega^2/c^2$ is the vacuum potential drop across the polar cap. In comparison, the radiation-reaction-limited particle Lorentz factor at the light cylinder is always much smaller than $\sigma_{\rm LC}$ in our simulations. It can be estimated by the following expression ({\em e.g.}, \citealt{2013ApJ...770..147C}):
\begin{equation}
\gamma^{\rm LC}_{\rm rad}=\sqrt{\frac{3e}{2f_{\rm rad}r^2_{\rm e}B_{\rm LC}}}.
\end{equation}
Here, $\gamma^{\rm LC}_{\rm rad}$ varies from $\approx 12$ for $\sigma_{\rm LC}=50$ to $\gamma^{\rm LC}_{\rm rad}\approx 38$ for $\sigma_{\rm LC}\approx 500$. We note that in Crab-like pulsars this quantity can be as high as $\gamma^{\rm LC}_{\rm rad}\sim 10^4-10^5$.

It is worth noticing here that in contrast to plane parallel reconnection studies at very high magnetization \citep{2014ApJ...783L..21S, 2016ApJ...816L...8W, 2014PhRvL.113o5005G}, the particle spectra do not converge to a broad, hard ($p\approx -1$) power law.  We see hints of such a hard power-law tail at intermediate radii only. Instead, all particles end up gaining a very similar amount of energy regardless of their histories. The characteristic energy is naturally set by $\sigma_{\rm LC}$ because it represents the total available magnetic energy per particle. The fact that the sheets become so close to each other (and eventually merging with each other) at large radii could be a possible explanation for this important difference with plane parallel studies where the sheet is isolated or as far away as possible from its neighbor thus preventing any interaction between them. This result is also consistent with the linear expansion of the sheet with radius found in the previous Section (Sect.~\ref{sect_delta}, Eqs.~\ref{eq_diss}-\ref{eq_kappa}). Assuming that the layer thickness scales with the mean particle Larmor radius in the sheet $\delta\sim \langle\gamma\rangle m_{\rm e} c^2/eB$ and with $\langle\gamma\rangle\approx \sigma_{\rm LC}\sim \phi_{\rm pc}/\Gamma_{\rm LC}\kappa_{\rm LC}$ (we dropped the factor 4) yields $\Delta\left(r\right)=\left(1/\pi\Gamma_{\rm LC}\kappa_{\rm LC}\right)\left(r/R_{\rm LC}\right)$, which coincides with the expression found in Eq.~(\ref{eq_kappa}). One can also recover this relation by setting the layer thickness to the local plasma skin-depth, $\delta\sim d_{\rm e}=(\langle\gamma\rangle m_{\rm e} c^2/4\pi n e^2)^{1/2}$.

\begin{figure}
\centering
\includegraphics[width=\hsize]{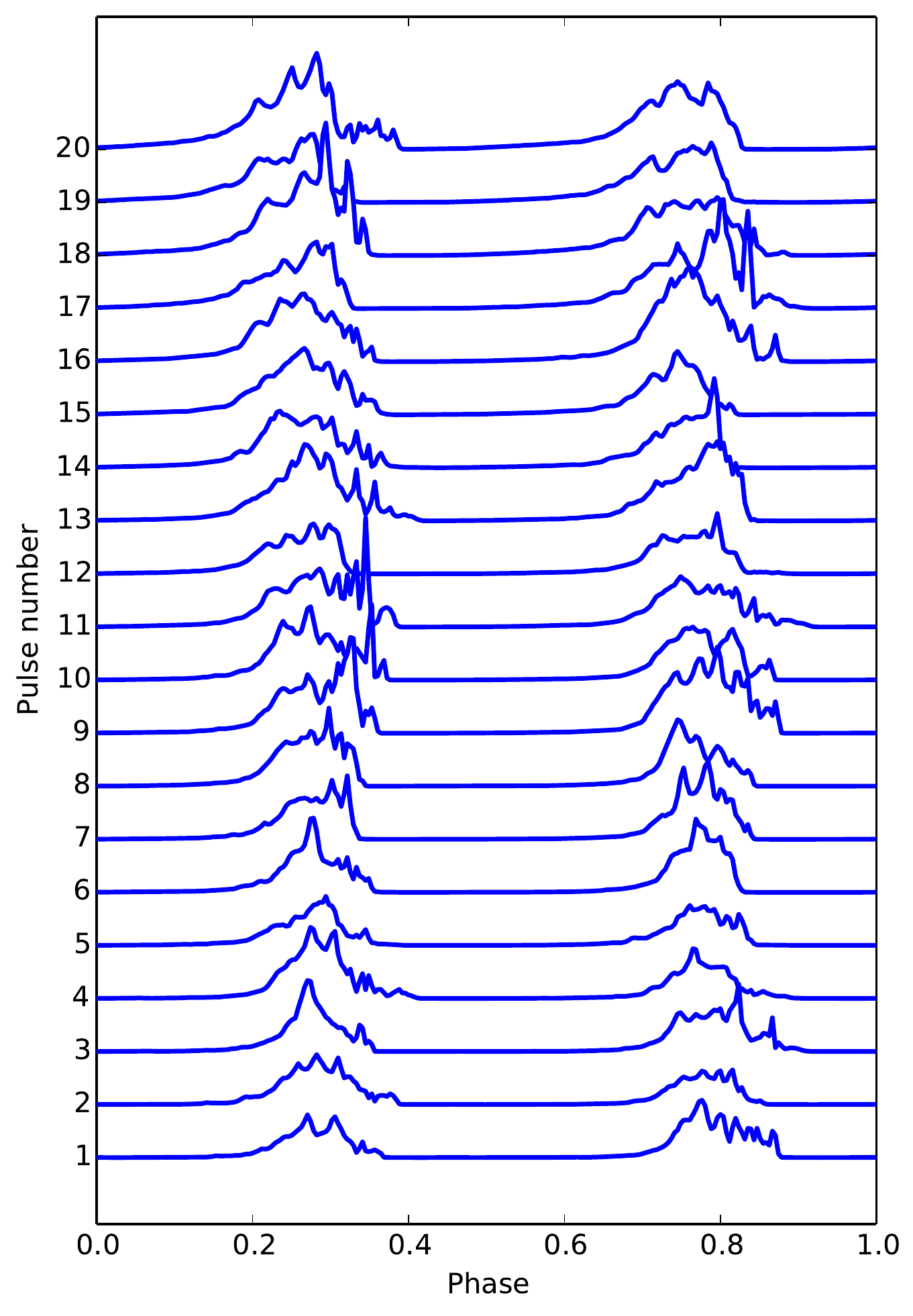}
\includegraphics[width=\hsize]{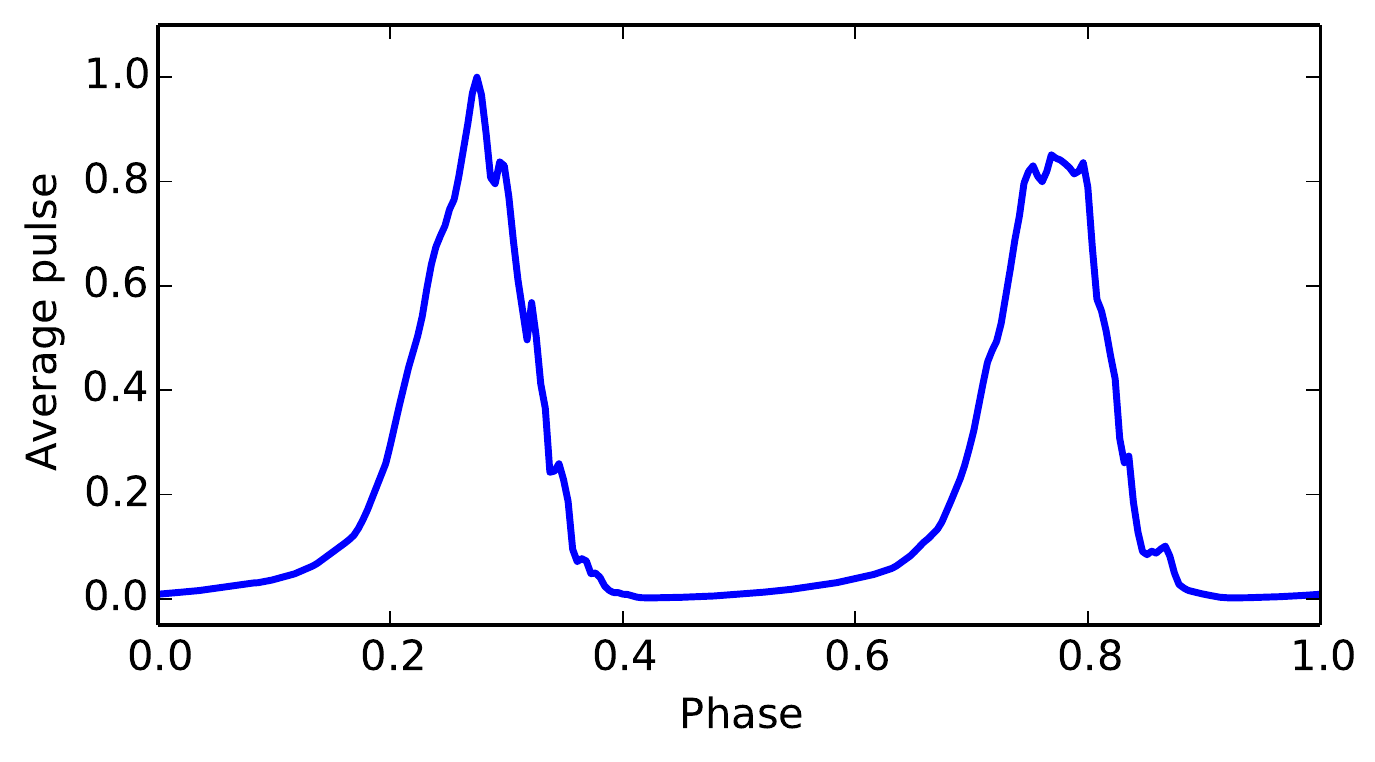}
\caption{Top: Series of individual pulse profiles computed after each pulsar period. Bottom: Pulse profile averaged over all periods. The origin of phases is defined by the direction of the north magnetic pole.}
\label{fig_pulse}
\end{figure}

The bottom panel in Figure~\ref{fig_spectra} shows the resulting spectral energy distribution ($\nu F_{\nu}$) averaged in azimuth as function of radius. Synchrotron radiation is the dominant process at the origin of the high-energy emission. In contrast with the particles, the mean synchrotron photon frequency decreases with radius (Fig.~\ref{fig_energy}). The highest-energy photons are found close to but outside of the light cylinder where there are both energetic particles and a strong magnetic field. The frequency-integrated flux is also concentrated at the light cylinder within a few $R_{\rm LC}$ (Fig.~\ref{fig_energy}, bottom panel). It is therefore sufficient to consider the innermost parts of the striped wind to model pulse profiles. Following \citet{2016MNRAS.457.2401C}, we compute the synchrotron flux as received by an observer at infinity looking in the plane of the simulation. Figure~\ref{fig_pulse} presents a series of individual light curves computed after each pulsar period and folded over the pulsar phase whose origin is defined by the direction of the north magnetic pole. The obtained light curves present two pulses at phase $0.25$ and $0.75$, in agreement with previous 3D simulations for an observer looking along the equator \citep{2016MNRAS.457.2401C, 2017arXiv170704323P}.

Thanks to the long integration time of the simulations, we can study the pulse-to-pulse variability. We find bright substructures randomly distributed within each pulse, reminiscent of micropulses observed in some pulsars in radio \citep{2006puas.book.....L}. These substructures are emitted by the particles confined within the plasmoids and shining towards the observer. Since most of the emission comes from the base of the striped wind, the synchrotron flux probes the most active reconnecting regions, that is, where there is the largest number of plasmoids and plasmoid mergers. However, due to the low photon flux at high energies, instruments are only sensitive to the average pulse profile. Averaging over all pulses washes out the intra-pulse variability and leaves the two main peaks which are very stable from one pulse to another (bottom panel in Fig.~\ref{fig_energy}).

%--------------------------------------------------------------------

\section{Conclusion}\label{sect_ccl}

We have investigated the formation and the dissipation of the striped pulsar wind structure using large 2D PIC simulations of a rotating split-monopole. We observe that shortly after its formation, the wind current sheet breaks up into a highly dynamical chain of magnetic islands separated by secondary current sheets, a configuration reminiscent of previous plane parallel reconnection studies. Close to the light cylinder, the dynamics of plasmoids is very time-dependent: they form, expand and merge with each other multiple times. This is the most dynamical, turbulent-like phase of reconnection which is accompanied by efficient dissipation of the Poynting flux \citep{2016ApJ...823...39Z}. At larger radii, adiabatic expansion takes over and no more plasmoids form or merge. Reconnection proceeds in a more laminar way and continues to dissipate the Poynting flux without any sign of saturation up to at least $150 R_{\rm LC}$ where the numerical box ends. The current sheet thickness expands linearly with radius in accordance with the increase of the plasma skin-depth in the wind.

Complete dissipation does not occur within the numerical domain (up to $90\%$) but simulations suggest that the dissipation radius is set by the plasma multiplicity at the light cylinder, $r_{\rm diss}/R_{\rm LC}=\pi\Gamma_{\rm LC} \kappa_{\rm LC}$. This corresponds to the location where the layer thickness is equal to the wavelength of the stripes. This remarkably simple result has important consequences. With multiplicities of order $\kappa\sim 10^3-10^4$ \citep{2001ApJ...554..624H, 2013MNRAS.429...20T} and $\Gamma_{\rm LC}\lesssim 100$, our results imply that the stripes should dissipate far before the wind reaches the termination shock in isolated systems, even in the Crab pulsar where $R_{\rm shock}/R_{\rm LC}\sim 10^9 \gg \Gamma_{\rm LC}\kappa\sim 10^5$-$10^6$ \citep{1990ApJ...349..538C, 1994ApJ...431..397M}. We note that this conclusion is not consistent with \citet{2001ApJ...547..437L} and \citet{2003ApJ...591..366K}. A possible explanation for this discrepancy could be that reconnection is very effective at dissipating magnetic energy in the inner zone of the striped region (about half is dissipated), where the wind is not super-Alfvenic. This transition region is not considered by the analytical models but it is well captured by the simulations. Another reason could be that we find no significant acceleration of the flow in the super-fast wind region as the stripes dissipate and hence no time-dilation effect to slow down the process in contrast to what \citet{2001ApJ...547..437L} found. This could be an artefact of the relatively low magnetization explored in our simulations.

What happens beyond the dissipation radius is unknown. What is clear, however, is that the current sheet will not lack electric charge, as long as there is enough charge at the light cylinder, because the particle drift velocities needed to carry the current are subluminal and decrease with radius \citep{2012SSRv..173..341A}. Magnetic dissipation results in efficient particle acceleration and emission of high-energy pulsed synchrotron emission. Even though magnetic dissipation is spread over hundreds of light-cylinder radii, the synchrotron flux piles up at the light cylinder. This implies that the pulsed high-energy emission probes the innermost parts of the pulsar wind where reconnection is most time-dependent, resulting in significant pulse-to-pulse variability \citep{2016MNRAS.457.2401C}. The particle spectra asymptotically converge to a narrow energy distribution with no sign of a high-energy power-law tail in contrast with plane parallel reconnection studies \citep{2014ApJ...783L..21S, 2014PhRvL.113o5005G, 2016ApJ...816L...8W}. The characteristic particle Lorentz factor in the wind after dissipation is set by the magnetization parameter at the light cylinder $\gamma\approx \sigma_{\rm LC}$ which also depends on the plasma multiplicity parameter $\sigma_{\rm LC}\propto \kappa^{-1}_{\rm LC}$ showing once again how critical this parameter is in this problem. It may be determined by both the plasma supply from the star {\em via} the polar-cap discharge and photon-photon annihilation within the current sheet \citep{1996A&A...311..172L, 2017arXiv170704323P}. While polar-cap cascade was extensively studied in the past, pair creation within the current sheet is still poorly understood. Dedicated studies of this phenomenon would be very valuable.

To conclude, the striped component of the pulsar wind is most likely fully annihilated far before it reaches the nebula. The wind is left with its aligned DC component (not present in this work) which may remain highly magnetized up to the termination shock. Three-dimensional simulations would be required to investigate the full latitude-dependent evolution of the striped wind. Note that even though the striped component disappears, a large-scale current sheet survives in the equatorial regions which might be a promising location for extra particle acceleration and magnetic dissipation in pulsar wind nebulae. Pulsars in binary systems such as gamma-ray binaries \citep{2013A&ARv..21...64D} or transitional millisecond pulsars \citep{2009Sci...324.1411A, 2014ApJ...790...39S} where the pulsar wind zone is truncated at much smaller radii than in isolated systems (by the companion star wind or by an accretion disk) may provide ideal environments to probe the innermost regions of the wind where we expect dissipation to be most active.

\begin{acknowledgements}
We thank A. Beloborodov, G. Dubus, A. Levinson, Y. Lyubarsky, E. Nakar, and A. Spitkovsky for discussions. BC acknowledges support from CNES and Labex OSUG@2020 (ANR10 LABX56). This research was also supported by the NASA Earth and Space Science Fellowship Program (grant NNX15AT50H to AP), Porter Ogden Jacobus Fellowship awarded by Princeton University to AP, NASA through Einstein Postdoctoral Fellowship grant number PF7-180165 awarded by the Chandra X-ray Center to AP, which is operated by the Smithsonian Astrophysical Observatory for NASA under contract NAS8-03060. The simulations presented in this article used computational resources of TGCC under the allocation t2016047669 made by GENCI, of the NASA/Ames HEC Program (SMD-16-7816), as well as NSF through an XSEDE computational time allocation TG-AST100035 on TACC Stampede.
\end{acknowledgements}

\bibliographystyle{aa}
\bibliography{striped}

\end{document}